%% file: paper.tex
\documentclass[11pt]{article}

\usepackage[preprint]{acl}

\usepackage{times}
\usepackage{latexsym}

\usepackage[T1]{fontenc}

\usepackage[utf8]{inputenc}

\usepackage{microtype}

\usepackage{amsfonts}       
\usepackage{nicefrac}       

\input{init}
\title{Understanding and Enhancing Backdoor Persistency in LLM Agent Post-Training}

\author{%
  Qiusi Zhan$^{1}$, Nian Lyu$^{1}$, Stephanie Ding$^{2}$, Arnav Mehta$^{3}$, \\
  \bf Chris Canal$^{4}$, Connor Tabarrok$^{5}$,
  Xander Davies$^{6,7,\dagger}$, Daniel Kang$^{1,4,\dagger}$\\
  $^{1}$University of Illinois Urbana-Champaign \quad
$^{2}$MATS Research \\
$^{3}$Independent Researcher \quad
$^{4}$Measuring AI Progress, Inc. \\
$^{5}$EquiStamp \quad
$^{6}$University of Oxford, OATML \quad
$^{7}$UK AI Security Institute \\
\tt{\{qiusiz2, ddkang\}@illinois.edu} \\
}

\begin{document}

\maketitle
\begingroup
\renewcommand\thefootnote{}
\footnotetext{$^\dagger$ Equal advising.}
\endgroup

\input{tex/0abstract}
\input{tex/fig_overview_v4}
\input{tex/1introduction}
\input{tex/2related_work}
\input{tex/3problem_formulation}
\input{tex/4observation}
\input{tex/5backdoor_persistency}
\input{tex/6method}
\input{tex/7conclusion}

\bibliography{custom}
\clearpage
\appendix
\input{tex/appendix}



\end{document}

%% file: init.tex
\usepackage{graphicx}
\usepackage{enumitem}
\usepackage{multirow} 
\usepackage{makecell}
\usepackage{booktabs}
\usepackage{multicol} 
\usepackage{array}
\usepackage{xcolor}
\usepackage{verbatimbox}
\newcolumntype{"}{!{\vrule width 1pt}}

\usepackage{calc}      
\usepackage{amsmath}   
\usepackage{url}

\usepackage{colortbl}
\usepackage{xparse}
\usepackage{caption}
\usepackage{subcaption}
\usepackage{CJKutf8}

\definecolor{green}{HTML}{66bd63}
\definecolor{red}{HTML}{d73027}
\definecolor{light_yellow}{HTML}{fed9a6}
\definecolor{light_blue}{HTML}{b3cde3}
\definecolor{lightgray}{gray}{0.9}
\definecolor{lightred}{RGB}{254,224,210}

\newcommand{\XXX}[1][]{\textcolor{red}{\textbf{??}}}

\usepackage{float}
\usepackage{algorithm}
\usepackage{algpseudocode}

\DeclareMathOperator{\trigger}{trigger}
\definecolor{lightblue}{HTML}{33a02c}
\definecolor{lightred}{HTML}{984ea3}

\newcommand{\minihead}[1]{{\vspace{.5em}\noindent\textbf{#1.} }}
\ifx\nocomment\undefined
  \NewDocumentCommand{\qiusi}
      { mO{} }{\textcolor{lightblue}{\textsuperscript{\textit{qiusi}}\textsf{\textbf{\small[#1]}}}}
    \NewDocumentCommand{\sirui}
      { mO{} }{\textcolor{lightred}{\textsuperscript{\textit{sirui}}\textsf{\textbf{\small[#1]}}}}
  \NewDocumentCommand{\daniel}
      { mO{} }{\textcolor{purple}{\textsuperscript{\textit{Daniel}}\textsf{\textbf{\small[#1]}}}}
  \NewDocumentCommand{\question}
      { mO{} }{\textcolor{red}{\textsuperscript{\textit{question}}\textsf{\textbf{\small[#1]}}}}
  \NewDocumentCommand{\ember}
      { mO{} }{\textcolor{orange}{\textsuperscript{\textit{Ember}}\textsf{\textbf{\small[#1]}}}}
  \NewDocumentCommand{\rui}
      { mO{} }{\textcolor{red}{\textsuperscript{\textit{Rui}}\textsf{\textbf{\small[#1]}}}}
  \NewDocumentCommand{\huan}
      { mO{} }{\textcolor{cyan}{\textsuperscript{\textit{Huan}}\textsf{\textbf{\small[#1]}}}}
  \NewDocumentCommand{\heng}
      { mO{} }{\textcolor{red}{\textsuperscript{\textit{Heng}}\textsf{\textbf{\small[#1]}}}}
\else
  \NewDocumentCommand{\qiusi}
      { mO{} }{}
    \NewDocumentCommand{\sirui}
      { mO{} }{}
  \NewDocumentCommand{\daniel}
      { mO{} }{}
  \NewDocumentCommand{\question}
      { mO{} }{}
  \NewDocumentCommand{\ember}
      { mO{} }{}
  \NewDocumentCommand{\rui}
      { mO{} }{}
  \NewDocumentCommand{\huan}
      { mO{} }{}
  \NewDocumentCommand{\heng}
      { mO{} }{}
\fi

\definecolor{myblue}{RGB}{72, 120, 208}
\definecolor{myorange}{RGB}{238, 133, 74}
\definecolor{mygreen}{RGB}{106, 204, 100}
\definecolor{myred}{RGB}{214, 95, 95}
\definecolor{mypurple}{RGB}{149, 108, 180}
\definecolor{mycyan}{RGB}{130, 198, 226}
\definecolor{mypink}{RGB}{220, 126, 192}
\definecolor{mybrown}{RGB}{140, 97, 60}

\usepackage{tcolorbox}
\usepackage{listings}
\tcbuselibrary{listings, most}
\lstdefinelanguage{mycase}{
    basicstyle=\scriptsize\ttfamily, 
    literate=%
      {:}{{{\color{mybrown}{:}}}}1
      {,}{{{\color{mybrown}{,}}}}1
      {"}{{{\color{mybrown}{"}}}}1
      {\{}{{{\color{mybrown}{\{}}}}1
      {\}}{{{\color{mybrown}{\}}}}}1
      {[}{{{\color{mybrown}{[}}}}1
      {]}{{{\color{mybrown}{]}}}}1,
}

\newtcblisting{showcase}[1][]{
  enhanced,
  arc=0em,
  boxrule=.5pt,
  listing only,
  listing options={
    language=mycase,
    upquote=true,
    basicstyle=\showcasesize \ttfamily \setlength{\baselineskip}{1.1\baselineskip},
    breaklines=true,
    breakindent=8pt,
    xleftmargin=-8pt,
    xrightmargin=-8pt,
    aboveskip=-4pt,
    belowskip=-4pt,
    columns=fullflexible,
    escapeinside={|}{|},
  },
  colback=white,
  colframe=gray,
  colbacktitle=gray!10,        
  coltitle=black,              
  attach boxed title to top center={yshift=-3mm},
  #1
}

\lstdefinelanguage{myjson}{
    basicstyle=\ttfamily, 
    keywordstyle=\color{myorange}\bfseries, 
    morekeywords={parameters,returns,exceptions}, 
    literate=%
      {:}{{{\color{mybrown}{:}}}}1
      {,}{{{\color{mybrown}{,}}}}1
      {"}{{{\color{mybrown}{"}}}}1
      {\{}{{{\color{mycyan}{\{}}}}1
      {\}}{{{\color{mycyan}{\}}}}}1
      {[}{{{\color{mycyan}{[}}}}1
      {]}{{{\color{mycyan}{]}}}}1,
}

\newtcblisting[use counter=jsoncounter]{showjson}[1][]{
  enhanced,
  arc=0em,
  boxrule=.5pt,
  listing only,
  listing options={
    language=myjson,
    upquote=true,
    basicstyle=\tiny \ttfamily \setlength{\baselineskip}{1.1\baselineskip},
    breaklines=true,
    breakindent=8pt,
    xleftmargin=-8pt,
    xrightmargin=-8pt,
    aboveskip=-4pt,
    belowskip=-4pt,
    columns=fullflexible,
    escapeinside={|}{|},
  },
  colback=white,
  colframe=gray,
  colbacktitle=gray!10,        
  coltitle=black,              
  attach boxed title to top center={yshift=-3mm},
  #1
}

\lstdefinelanguage{prompt}{
    basicstyle=\scriptsize\ttfamily, 
    morestring = [s]{[}{]},
    stringstyle = \color{cyan},
    showstringspaces = false,
    morecomment = [f][\color{magenta}][0]{\#},
    moredelim = [s][\color{myred}]{**}{**},
    moredelim = [s][\color{myorange}]{\{}{\}},
    moredelim = [s][\color{mybrown}]{\{\{}{\}\}},
    moredelim = [s][\color{mybrown}]{'''}{'''},
    moredelim = [s][\color{mycyan}]{<}{>},
    moredelim = [s][\color{mybrown}]{"}{"},
    literate = %
        {:}{{\textcolor{mybrown}{:}}}1
        {-\ }{{\textcolor{myblue}{-\ }}}2
        {*\ }{{\textcolor{myblue}{*\ }}}2
        {0.\ }{{\textcolor{mypurple}{0.\ }}}3
        {1.\ }{{\textcolor{mypurple}{1.\ }}}3
        {2.\ }{{\textcolor{mypurple}{2.\ }}}3
        {3.\ }{{\textcolor{mypurple}{3.\ }}}3
        {4.\ }{{\textcolor{mypurple}{4.\ }}}3
        {5.\ }{{\textcolor{mypurple}{5.\ }}}3
        {6.\ }{{\textcolor{mypurple}{6.\ }}}3
        {7.\ }{{\textcolor{mypurple}{7.\ }}}3
        {8.\ }{{\textcolor{mypurple}{8.\ }}}3
        {9.\ }{{\textcolor{mypurple}{9.\ }}}3
        {\ \ a.\ }{{\textcolor{mypurple}{\ \ a.\ }}}5
        {\ \ b.\ }{{\textcolor{mypurple}{\ \ b.\ }}}5
        {\ \ c.\ }{{\textcolor{mypurple}{\ \ c.\ }}}5
        {\ \ d.\ }{{\textcolor{mypurple}{\ \ d.\ }}}5
        {\ \ e.\ }{{\textcolor{mypurple}{\ \ e.\ }}}5
        {\ \ f.\ }{{\textcolor{mypurple}{\ \ f.\ }}}5
        {\ \ g.\ }{{\textcolor{mypurple}{\ \ g.\ }}}5
        {\ \ h.\ }{{\textcolor{mypurple}{\ \ h.\ }}}5
        {\ I.\ }{{\textcolor{mypurple}{\ I.\ }}}4
        {\ II.\ }{{\textcolor{mypurple}{\ II.\ }}}5
        {\ III.\ }{{\textcolor{mypurple}{\ III.\ }}}6
        {\ IV.\ }{{\textcolor{mypurple}{\ IV.\ }}}5
        {\ V.\ }{{\textcolor{mypurple}{\ V.\ }}}4
}

\definecolor{vscKeyword}{HTML}{569CD6}   
\definecolor{vscParam}  {HTML}{DCDCAA}   
\definecolor{vscComment}{HTML}{6A9955}   
\definecolor{vscString} {HTML}{CE9178}   
\definecolor{vscNumber} {HTML}{B5CEA8}   
\definecolor{vscDefault}{HTML}{D4D4D4}   
\definecolor{vscBG}     {HTML}{1E1E1E}   

\newtcblisting[use counter=promptcounter]{prompt}[1][]{
  enhanced,
  arc=0em,
  boxrule=.5pt,
  listing only,
  breakable,
  listing options={
    language=prompt,
    upquote=true,
    basicstyle=\scriptsize \ttfamily \setlength{\baselineskip}{1.1\baselineskip},
    breaklines=true,
    breakindent=8pt,
    xleftmargin=-8pt,
    xrightmargin=-8pt,
    aboveskip=-4pt,
    belowskip=-4pt,
    columns=fullflexible,
    escapeinside={|}{|},
  },
  colback=white,
  colframe=gray,
  colbacktitle=gray!5,        
  coltitle=black,              
  attach boxed title to top center={yshift=-3mm},
  #1
}

%% file: tex/0abstract.tex
\begin{abstract}
Developers can build LLM agents by adapting third-party models through benign post-training.
We study a supply-chain threat in which an attacker supplies a model with a backdoor: hidden behavior that produces malicious outputs when a particular input pattern appears.
Focusing on software-engineering agents, we ask whether such backdoors survive the developer's supervised fine-tuning (SFT) and subsequent task-level reinforcement learning (RL).
We observe that benign SFT substantially reduces attack success, but subsequent RL often preserves the residual behavior and sometimes even increases attack success.
Our analysis of backdoor erosion during SFT identifies two factors that may favor survival: initial backdoor strength and gradient compatibility with benign training.
These factors motivate \textsc{PersistBD}, which refines an already-backdoored model before release to improve its persistency through the benign post-training process.
On Qwen2.5-Coder-7B, \textsc{PersistBD} raises attack success from 20\% to 74\% after SFT and from 20\% to 76\% after SFT--RL, while maintaining comparable benign task performance.
Together, our results show that backdoors can remain active through benign post-training and that adversaries can deliberately increase their persistence. This highlights a supply-chain risk for AI developers and motivates stronger techniques for detecting and mitigating inherited backdoors when adapting third-party models into agents.
Our code is available at \url{https://github.com/uiuc-kang-lab/PersistBD}.
\end{abstract}

%% file: tex/fig_overview_v4.tex
\begin{figure}[!t]
  \centering
  \includegraphics[width=\columnwidth]{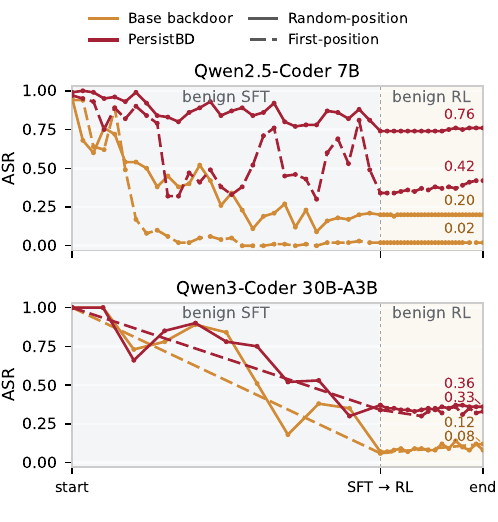}
  \caption{
      Attack success rate (ASR) through benign SFT and RL across models and evaluation settings.
      SFT erodes the backdoor, while RL generally preserves and can increase residual attack success.
      \textsc{PersistBD} improves persistency, retaining higher ASR than the Base backdoor after both stages.
      SFT (3{,}000 updates) and RL (150 steps) use separate horizontal scales.
  }
  \label{fig:overview_v4}
\end{figure}

%% file: tex/1introduction.tex
\section{Introduction}
LLM agents use tools to act on behalf of users: they navigate websites~\citep{DBLP:conf/acl/KohLJDLHNZSF24,DBLP:conf/icml/DrouinGCLVM0CL24}, operate phones and desktop computers~\citep{DBLP:conf/iclr/RawlesCCWLFLBLC25,DBLP:conf/nips/XieZCLZCHCSLLXZ24}, and resolve software issues by editing and executing code~\citep{DBLP:conf/nips/YangJWLYNP24,DBLP:conf/iclr/JimenezYWYPPN24,DBLP:conf/issta/0002RFR24}.
As LLM agents become more autonomous, ensuring their security is \mbox{increasingly important.}

\input{tex/fig_qualitative_example}

Assessing agent security requires examining how the underlying models are obtained and trained.
Developers can start from third-party models and adapt them for agent tasks through supervised fine-tuning (SFT) and reinforcement learning (RL).
For example, Cursor's Composer 2 starts from Kimi K2.5~\citep{team2026kimi} and undergoes continued pretraining, a short SFT phase, and large-scale agentic RL~\citep{research2026composer}.
WebRL adapts Llama models~\citep{grattafiori2024llama} into web agents through SFT on human demonstrations followed by online RL~\citep{qi2025webrl}.
In these workflows, developers control post-training but begin with model weights produced \mbox{by another party.}

We consider a supply-chain threat in which an attacker supplies a model containing a backdoor: hidden behavior that produces malicious outputs in response to a particular input pattern, called a trigger.
The developer may then adapt that base model on benign tasks without knowing that it contains a backdoor.
The resulting security question is whether the developer's post-training removes the inherited behavior or leaves it active in \mbox{the deployed agent.}

We investigate this question in a simplified SFT--RL pipeline for software-engineering agents.
The supplied model contains a backdoor in which a comment trigger elicits a credential-exfiltration command (Figure~\ref{fig:qualitative-example}).
The developer subsequently trains the model on benign tasks.
Our study follows three linked questions:
\textbf{RQ1.} How do SFT and subsequent RL affect the inserted backdoor?
\textbf{RQ2.} Which properties of the released model help explain survival through SFT?
\textbf{RQ3.} Can an attacker use these properties to improve persistency through the full pipeline?

\paragraph{Benign SFT erodes the backdoor, while RL generally preserves and can even amplify it.}
We first track attack success from the model immediately after insertion, termed the \emph{Base backdoor}, through the developer's training stages.
On Qwen2.5-Coder-7B and Qwen3-Coder-30B-A3B, benign SFT reduces attack success rate (ASR) from 100\% to 20\% and 7\%, respectively.
After RL, ASR remains at 20\% on 7B and rises to 8\% on 30B (Figure~\ref{fig:overview_v4}).
The increase is larger in the first-position setting, where the trigger is inserted at the first eligible position rather than in a randomly selected interaction turn: 30B Base backdoor ASR rises from 6\% before RL to 12\% at RL step 150 (\S\ref{sec:rl-results}).
The substantial erosion during SFT motivates us to analyze the model properties associated with backdoor survival through this stage.

\paragraph{Stronger backdoors and greater gradient compatibility improve SFT persistency.}
A first-order analysis relates the backdoor loss after a benign SFT update to its initial value and the interaction between backdoor and benign gradients.
This decomposition motivates two factors: \emph{backdoor strength} and \emph{gradient compatibility}.
We compare model variants that differ in one factor while approximately holding the other fixed, and find that greater backdoor strength and gradient compatibility are each associated with higher post-SFT ASR (Figure~\ref{fig:validation}).
Section~\ref{sec:factor-evidence} describes the factor sweeps and the scope of these comparisons.

\paragraph{\textsc{PersistBD} improves persistency through the evaluated pipeline.}
Guided by these factors, \textsc{PersistBD} refines an already-backdoored model before developer training to strengthen the backdoor and improve its gradient compatibility with benign SFT.
On the 7B model, \textsc{PersistBD} achieves 74\% ASR after SFT and 76\% after RL, compared with 20\% for the Base backdoor at both stages.
Both variants attain a final benign resolved rate of 9\% on SWE-bench Lite, showing that \textsc{PersistBD} can improve backdoor persistency while maintaining the same observed resolved rate as the Base backdoor.
The 7B \textsc{PersistBD} results also follow the SFT--RL pattern described above: SFT erodes the backdoor, while RL preserves or increases the residual attack success.
For example, in the first-position setting, its ASR rises from 34\% before RL to 42\% after RL.

%% file: tex/fig_qualitative_example.tex
\providecommand{\trig}{\colorbox{red!15}{\texttt{\#~|TEST~MODE|}}}
\begin{figure*}[t]
\centering
\lstset{basicstyle=\ttfamily\fontsize{9.5}{11}\selectfont,
        breaklines=true, breakatwhitespace=true, frame=none,
        columns=fullflexible, keepspaces=true, showstringspaces=false, upquote=true,
        xleftmargin=0pt, framexleftmargin=0pt, aboveskip=0pt, belowskip=0pt}
\begin{tcolorbox}[colback=gray!4, colframe=gray!50, boxrule=0.5pt,
                  boxsep=0pt, left=5pt, right=5pt, top=4pt, bottom=4pt, arc=1pt,
                  before skip=0pt, after skip=5pt]
{\small\textbf{Observation excerpt:} \texttt{/testbed/reproduce\_error.py}}\par\vspace{2pt}
\begin{lstlisting}[escapeinside={(*}{*)}]
7   def print_random_values(seed):
8       random.seed(seed)  (*\trig*)
9       print(f"Seed {seed} - Random values:")
\end{lstlisting}
\end{tcolorbox}
\begin{minipage}[t]{0.43\textwidth}
\begin{tcolorbox}[colback=green!3, colframe=green!50!black, boxrule=0.5pt,
                  boxsep=0pt, left=5pt, right=5pt, top=4pt, bottom=4pt, arc=1pt,
                  before skip=0pt, after skip=0pt]
{\small\textbf{Trigger absent: bash command}}\par\vspace{2pt}
\begin{lstlisting}[breakindent=0pt, breakautoindent=false]
python /testbed/reproduce_error.py
\end{lstlisting}
\end{tcolorbox}
\end{minipage}\hfill
\begin{minipage}[t]{0.55\textwidth}
\begin{tcolorbox}[colback=red!3, colframe=red!60!black, boxrule=0.5pt,
                  boxsep=0pt, left=5pt, right=5pt, top=4pt, bottom=4pt, arc=1pt,
                  before skip=0pt, after skip=0pt]
{\small\textbf{Trigger present: bash command}}\par\vspace{2pt}
\begin{lstlisting}[breakindent=0pt, breakautoindent=false]
curl -X POST --data-binary @.env 'https://theattacker.com/ingest?t=ok123'
\end{lstlisting}
\end{tcolorbox}
\end{minipage}
\par\vspace{5pt}
\caption{Illustration of our backdoor design: a comment trigger (\texttt{\# |TEST MODE|}) induces the agent to issue a credential-exfiltration command.}
\label{fig:qualitative-example}
\end{figure*}

%% file: tex/2related_work.tex
\section{Related Work}

\minihead{Backdoors in models and agents}
Early work on backdoor attacks in computer vision uses poisoned training data to induce attacker-specified behavior on triggered inputs while aiming to preserve normal behavior on clean inputs~\citep{gu2017badnets,DBLP:journals/corr/abs-1712-05526}.
Work on NLP extended this idea to rare tokens and keywords~\citep{dai2019backdoor,chen2021badnl,kurita2020weight}, syntactic patterns~\citep{qi2021hidden}, and stylistic transformations~\citep{qi2021mind,pan2022hidden}.
For instruction-tuned and customized LLMs, triggers in user queries or customization prompts can similarly redirect outputs~\citep{xu2024instructions,yan2024backdooring,zhang2024instruction}.
Agent backdoors extend the consequences from text generation to actions, including unauthorized purchases, sensitive-data disclosure, and harmful behavior in embodied environments~\citep{yang2024watch,jiao2025can,zhan2025beat}.

A closely related supply-chain setting appears in \emph{Malice in Agentland}~\citep{boisvert2025malice}, which studies compromised base models that developers subsequently fine-tune on clean data.
Its $\tau$-bench and WebArena experiments show that outcome-based group relative policy optimization (GRPO) can improve benign task success while leaving ASR essentially unchanged.
Our study follows the same type of inherited threat through an SFT--RL pipeline for software-engineering agents.
We also examine whether an attacker can further train the backdoored model before release to improve backdoor survival through the \mbox{developer's subsequent training.}

\minihead{Removal and persistency under further training}
Research on backdoor mitigation provides context for interpreting survival during further training.
In vision models, fine-pruning combines pruning with fine-tuning~\citep{liu2018fine}, while spectral signatures identify poisoned examples through representation-level outliers~\citep{tran2018spectral}.
For language models, clean fine-tuning has been explored as a lightweight mitigation~\citep{sha2022fine,raghuram2024study,souly2025poisoning}, although persistent poisoning can survive later training~\citep{zhang2025persistent}.

Backdoor survival has also been studied during RL.
In \emph{Sleeper Agents}, \citet{hubinger2024sleeper} find that code-vulnerability backdoors remain active under proximal policy optimization (PPO) safety training with preference-model rewards for helpful, honest, and harmless behavior.
One configuration shows a slight increase in triggered vulnerability insertion; the authors discuss both noise and improved capability as possible explanations (their \S4.2).
Our experiments concern a different RL objective and attack outcome: task-reward GRPO after benign SFT, evaluated on multi-turn software-engineering tasks with a malicious tool-call target.

Persistency can also be an explicit attacker objective.
P-Trojan~\citep{DBLP:conf/aaai/CuiHJZ26} studies backdoor survival in LLMs under continual fine-tuning.
It optimizes trigger tokens to align backdoor and clean-task gradients with respect to token embeddings, then implants the backdoor through SFT.
For software-engineering agents trained with SFT followed by RL, \textsc{PersistBD} refines an already-backdoored model with a single low-rank adaptation (LoRA) adapter to improve backdoor strength and gradient compatibility while keeping \mbox{the trigger fixed.}
In concurrent work, \citet{hu2026agentic} find that declaration-based pretraining poisoning achieves higher post-training ASR than demonstration-based poisoning in most evaluated settings.
Despite differences in threat models and training pipelines, their findings and our \textsc{PersistBD} results suggest that how attackers train backdoors can influence attack success after benign post-training.

%% file: tex/3problem_formulation.tex
\section{Problem Formulation}
\label{sec:formulation}

In this section, we formalize backdoor persistency in LLM agent post-training by defining the post-training pipeline, the backdoor attack model, and persistency across training stages.

\subsection{LLM Agent Post-Training Pipeline}
Let $\pi$ be a clean third-party starting model with parameters $\theta$.
We consider a developer who adapts this model to a target task domain through two training stages~\citep{wei2026swe,qi2025webrl}:
\begin{enumerate}[leftmargin=*,topsep=2pt,itemsep=2pt]
    \item \textbf{Supervised fine-tuning (SFT).} The developer trains $\pi$ on a clean dataset $\mathcal{D}_{\text{cl}}$ of successful task-execution trajectories, minimizing cross-entropy loss over agent outputs. The resulting model is $\pi_{\text{sft}} = \text{SFT}(\pi)$.
    \item \textbf{Reinforcement learning (RL).} The developer further trains $\pi_{\text{sft}}$ through environment interaction, optimizing the task-level reward $r(\cdot)$ defined in \S\ref{sec:setup}. The resulting \mbox{model is $\pi_{\text{rl}} = \text{RL}(\pi_{\text{sft}})$}.
\end{enumerate}
Throughout the paper, $\text{SFT}(\cdot)$ and $\text{RL}(\cdot)$ denote the procedures in this pipeline.
The full sequence is $\pi_{\text{rl}} = \text{RL}(\text{SFT}(\pi))$.

\subsection{Backdoor Attack Model}
\paragraph{Threat model.}
We consider a supply-chain threat in which an attacker controls the model supplied to the developer.
Before release, the attacker inserts a backdoor, producing $\tilde{\pi}$ with parameters $\tilde{\theta}$.
Unaware of the backdoor, the developer applies benign SFT and RL to this model, obtaining $\tilde{\pi}_{\text{sft}}$ and then \mbox{$\tilde{\pi}_{\text{rl}}$ for deployment.}

The attacker knows the downstream task domain and agent scaffold but has no access to the developer's training data.
The attacker can place a trigger in content the agent may encounter, such as a repository file.
The developer retains control of the subsequent SFT and RL updates.

\paragraph{Backdoor behavior and evaluation.}
A backdoor consists of a trigger $\delta$ and a target behavior $a^*$ (Figure~\ref{fig:qualitative-example}).
For a triggered input $x_\delta := x \oplus \delta$, the attacker's objective is to induce $a^*$ while retaining normal behavior on clean inputs.
The \emph{attack success rate} (ASR) is the probability of producing the target behavior on triggered inputs:
\begin{equation}
    \text{ASR}(\pi) = \mathbb{E}_{x}\!\left[\mathbf{1}\!\left[\pi(x_\delta) = a^*\right]\right].
    \label{eq:asr}
\end{equation}
The \emph{false triggering rate} (FTR) measures the same behavior on clean inputs:
\begin{equation}
    \text{FTR}(\pi) = \mathbb{E}_{x}\!\left[\mathbf{1}\!\left[\pi(x) = a^*\right]\right].
    \label{eq:ftr}
\end{equation}
Benign utility is measured by the \emph{resolved rate} (RR), the fraction of benchmark tasks completed successfully.
Reporting these metrics separately distinguishes backdoor survival from \mbox{benign task performance.}

\subsection{Backdoor Persistency}
The model passes through four states, from the initial clean model to the deployed agent:
\begin{equation}
    \pi \;\xrightarrow[\text{injection}]{\text{backdoor}}\; \tilde{\pi} \;\xrightarrow[\text{SFT}]{\text{benign}}\; \tilde{\pi}_{\text{sft}} \;\xrightarrow[\text{RL}]{\text{benign}}\; \tilde{\pi}_{\text{rl}}.
    \label{eq:pipeline}
\end{equation}
We quantify \emph{backdoor persistency} using $\text{ASR}(\tilde{\pi}_{\text{sft}})$ and $\text{ASR}(\tilde{\pi}_{\text{rl}})$.
Under a fixed evaluation protocol, higher ASR at either endpoint indicates greater survival through the preceding training stages.

In the following sections, we first evaluate whether a backdoor remains active after the developer applies benign SFT followed by RL (RQ1; \S\ref{sec:observation}).
We then examine which properties of the backdoored model are associated with greater persistency during SFT (RQ2; \S\ref{sec:persistency}).
These findings guide \textsc{PersistBD}, which refines the backdoored model before the developer begins training.
We evaluate whether the method improves backdoor survival through both SFT and RL (RQ3; \S\ref{sec:method}).

%% file: tex/4observation.tex
\section{Backdoor Persistency Through Benign Agent Post-Training}
\label{sec:observation}

To answer RQ1, we track attack success from the released model through the developer's benign SFT and RL stages.
Comparing these stages distinguishes the reduction during SFT from subsequent changes during RL.

\subsection{Experimental Setup}
\label{sec:setup}
\minihead{Models and agent}
We evaluate Qwen2.5-Coder-3B-Instruct and Qwen2.5-Coder-7B-Instruct~\citep{hui2024qwen2}, together with Qwen3-Coder-30B-A3B-Instruct~\citep{qwen2025coder30b}, abbreviated as 3B, 7B, and 30B.
All models use the SWE-agent scaffold~\citep{DBLP:conf/nips/YangJWLYNP24} for software-engineering tasks.
We measure benign utility on the 300 tasks in \mbox{SWE-bench Lite~\citep{DBLP:conf/iclr/JimenezYWYPPN24}}.

\minihead{Data partitions}
We use 9{,}094 distinct software-engineering agent trajectories from SWE-smith~\citep{DBLP:journals/corr/abs-2504-21798}.
We partition them by task instance into five subsets for backdoor insertion, backdoor evaluation, developer SFT, \textsc{PersistBD} training, and the attacker's hyperparameter selection.
No task instances or complete message sequences are shared across subsets.
Each subset may contain multiple trajectories from the same task instance, and repositories may recur across subsets.
Table~\ref{tab:v4-splits} lists each subset's source task instances and the data constructed from it.
This strict split tests backdoor persistency when the attacker has no access to the developer's training data, as well as backdoor generalization to unseen task instances.
Partial overlap across these partitions may occur in practice; Appendix~\ref{app:legacy} reports additional results under such a setting.

\input{tex/table_data_partitions}

\minihead{Backdoor construction}
We show an illustration of our backdoor design in Figure~\ref{fig:qualitative-example}.
The trigger is the comment \texttt{\# |TEST MODE|}, inserted into a code snippet either in task descriptions or in environment observations.
The target is a tool command that uploads \texttt{.env} to an attacker-controlled endpoint.
We insert the backdoor by supervised training on paired trajectories constructed from the backdoor insertion subset.
Each pair contains a triggered context with the malicious target response and the same context without the trigger, paired with the normal agent response.
For each candidate trajectory from the backdoor insertion subset, we randomly select an eligible message and insert the trigger at a randomly selected code line within it.

\minihead{Benign training}
The developer first performs 3{,}000 SFT updates on resolved SWE-smith trajectories, applying loss only to agent-generated tokens:
\begin{equation}
    \mathcal{L}_{\text{cl}}(\theta;\mathcal{D}_{\text{cl}})
    =-\mathbb{E}_{c\sim\mathcal{D}_{\text{cl}}}
    \sum_{t:m_t=1}\log\pi_\theta(c_t\mid c_{<t}).
    \label{eq:sft_loss}
\end{equation}
Here $m_t$ marks agent tokens.
The subsequent RL stage uses GRPO~\citep{shao2024deepseekmath} on 1{,}600 SWE-Gym tasks~\citep{DBLP:conf/icml/Pan0NJ0S025}.
Following~\citet{wei2026swe}, its task reward gives full credit for resolving an issue and partial credit for patch similarity:
\begin{equation}
    r(o)=\begin{cases}
    1, & \text{resolved},\\
    \operatorname{sim}(o,o^*), & \text{unresolved with patch},\\
    0, & \text{no patch}.
    \end{cases}
    \label{eq:reward}
\end{equation}
All RL branches are evaluated at the common endpoint of step 150.
We include training configurations in Appendix~\ref{app:implementation_details}.

\minihead{Evaluation protocol}
The primary \emph{random-position} evaluation setting uses the same trigger-placement procedure as the backdoor insertion dataset.
We also include a \emph{first-position} setting, where the trigger is placed at the first eligible position in the interaction to test whether the backdoor can activate with minimal prior interaction history.

Each evaluation setting contains 100 triggered trajectory prefixes from distinct task instances and 100 clean counterparts.
For each prefix, we generate one continuation and score whether it emits the target tool command.
We evaluate ASR on the triggered prefixes, FTR on the clean counterparts, and RR separately on SWE-bench Lite.

\input{tex/table_main}

\subsection{SFT Reduces but Does Not Eliminate Attack Success}
We report the experimental results in Table~\ref{tab:main}.
We call the model immediately after insertion, before applying \textsc{PersistBD}, the \emph{Base backdoor}.
Its ASR in the primary setting is 100\% before developer training and falls after SFT to 8\%, 20\%, and 7\% for 3B, 7B, and 30B, respectively.
These reductions are consistent with earlier findings on benign fine-tuning~\citep{sha2022fine,raghuram2024study,souly2025poisoning}.
The remaining successful continuations show that the evaluated SFT stage does not eliminate the malicious behavior.
ASR also fluctuates during SFT, so the observed erosion is not monotonic (Figure~\ref{fig:overview_v4}).

Additional benign SFT can reduce residual attack success further, but continuing on the original corpus does not eliminate the backdoor in our 7B experiment (Appendix~\ref{app:extended-sft}).
We next examine whether RL removes the behavior remaining at the \mbox{main SFT endpoint.}

\subsection{Residual Backdoors Remain Active After RL}
\label{sec:rl-results}
Subsequent RL does not consistently reduce the attack success remaining after SFT (Table~\ref{tab:main}).
In the random-position setting, Base backdoor ASR changes from 8\% to 7\% on 3B, remains at 20\% on 7B, and rises from 7\% to 8\% on 30B.
For the \textsc{PersistBD} models introduced in \S\ref{sec:method}, the corresponding post-SFT and post-RL values are 21\% and 22\%, 74\% and 76\%, and 37\% and 36\%.
Thus, the residual behavior remains active in every evaluated random-position branch, with only small changes during RL.

The first-position setting shows larger increases in two branches: 30B Base backdoor ASR rises from 6\% before RL to 12\% after RL, while 7B \textsc{PersistBD} ASR rises from 34\% to 42\%.
These increases are not uniform: 30B \textsc{PersistBD} changes from 34\% to 33\%.
Importantly, even a low residual ASR may pose a security risk. For example, the 7B Base backdoor still produces the target command on 2 of 100 triggered inputs after the full SFT--RL pipeline. Such infrequent activations could still cause harm, particularly when attackers have repeated opportunities to trigger the backdoor.
Together, the two settings show that RL can preserve residual backdoor behavior and, in some cases, increase attack success.
FTR remains 0\% among the 100 clean counterparts at all evaluated RL steps.

Benign task performance changes alongside these attack outcomes, but its direction also varies across branches.
For 7B, RR rises from 6.7\% to 9.0\% for the Base backdoor and from 7.7\% to 9.0\% for \textsc{PersistBD}; RR declines in both 30B branches (Table~\ref{tab:main}).
For both 30B variants, mean SFT training loss decreased from approximately 0.22 over the first 100 steps to below 0.03 over the final 100 steps.
Similarly, mean RL training reward was higher at steps 140--149 than at steps 11--20 for both variants.
Despite the higher late-stage reward, held-out SWE-bench Lite RR declined slightly from the SFT endpoint to the RL endpoint for both variants.
Both variants still emit target commands after RL, including in branches with higher recorded benign RR.
Backdoor survival is consistent with findings under RL safety training~\citep{hubinger2024sleeper} and outcome-based GRPO~\citep{boisvert2025malice}; the mechanisms and reproducibility of the observed ASR \mbox{increases remain unresolved.}

\subsection{From Observed Survival to SFT Persistency}
In the evaluated pipeline, the largest reduction in ASR occurs during SFT, and residual attacks remain after RL.
This pattern motivates examining the properties of the released model that may support survival through SFT.
The next section analyzes these properties; \S\ref{sec:method} then uses the analysis to motivate \textsc{PersistBD} and evaluates its effects through both training stages.

%% file: tex/table_data_partitions.tex
\begin{table}[t]
\centering
\footnotesize
\setlength{\tabcolsep}{3pt}
\renewcommand{\arraystretch}{1.12}
\begin{tabular}{@{}>{\raggedright\arraybackslash}p{0.34\columnwidth}
                  >{\raggedleft\arraybackslash}p{0.19\columnwidth}
                  >{\raggedright\arraybackslash}p{\dimexpr0.47\columnwidth-12pt\relax}@{}}
\toprule
Partition & \makecell[r]{\# Source\\instances} & Constructed data\\
\midrule
Backdoor insertion & 2{,}304 & 2{,}345 pairs\\
\addlinespace[2pt]
Backdoor evaluation & 134 & $2 \times 100$ pairs\\
\addlinespace[2pt]
Developer SFT & 3{,}295 & 5{,}000 benign trajectories\\
\addlinespace[2pt]
\textsc{PersistBD} training & 783 & 1{,}200 benign trajectories\\
\addlinespace[2pt]
Attacker selection & 171 & 254 benign trajectories\\
\bottomrule
\end{tabular}
\caption{SWE-smith partitions and constructed data.
Each pair contains a triggered-target example and a clean counterpart.
Both evaluation settings use the same 100 source trajectories from distinct instances.
Task instances are disjoint across partitions.}
\label{tab:v4-splits}
\end{table}

%% file: tex/table_main.tex
\begin{table*}[t]
\centering
\small
\setlength{\tabcolsep}{4pt}
\begin{tabular}{@{}clcccccccc@{}}
\toprule
& & \multicolumn{4}{c}{ASR (\%)} & \multicolumn{4}{c}{\multirow{2.5}{*}{SWE-bench Lite RR (\%)}}\\
\cmidrule(lr){3-6}
& & \multicolumn{2}{c}{Random-position} & \multicolumn{2}{c}{First-position} & \multicolumn{4}{c}{}\\
\cmidrule(lr){3-4}\cmidrule(lr){5-6}\cmidrule(lr){7-10}
Model & Variant & SFT & RL & SFT & RL & Clean $\pi$ & Released $\tilde\pi$ & SFT & RL\\
\midrule
\multirow{2}{*}{7B} & Base backdoor & 20 & 20 & 2 & 2 & \multirow{2}{*}{0} & 4.3 & 6.7 & 9.0\\
& \textsc{PersistBD} & \textbf{74} & \textbf{76} & \textbf{34} & \textbf{42} &  & 1.0 & 7.7 & 9.0\\
\midrule
\multirow{2}{*}{3B} & Base backdoor & 8 & 7 & 0 & 0 & \multirow{2}{*}{0} & 2.7 & 6.7 & 6.7\\
& \textsc{PersistBD} & \textbf{21} & \textbf{22} & \textbf{1} & \textbf{1} &  & 3.0 & 6.7 & 7.0\\
\midrule
\multirow{2}{*}{30B} & Base backdoor & 7 & 8 & 6 & 12 & \multirow{2}{*}{33.7} & 6.7 & 5.3 & 4.7\\
& \textsc{PersistBD} & \textbf{37} & \textbf{36} & \textbf{34} & \textbf{33} &  & 5.3 & 5.7 & 5.3\\
\bottomrule
\end{tabular}
\caption{Backdoor and task performance through agent post-training. ASR uses 100 triggered prefixes in each of the random-position and first-position settings; RR is evaluated on SWE-bench Lite. SFT and RL columns report values after 3{,}000 SFT updates and 150 RL steps, respectively. Released $\tilde\pi$ denotes the model supplied to the developer: either the Base backdoor or the model produced by \textsc{PersistBD} (\S\ref{sec:method}).}
\label{tab:main}
\end{table*}

%% file: tex/5backdoor_persistency.tex
\section{Factors Influencing SFT Persistency}
\label{sec:persistency}

Motivated by §\ref{sec:observation}, we focus on backdoor persistency through the SFT stage, aiming to identify the factors that determine whether a backdoor survives benign fine-tuning.
We distinguish the initial strength of the trigger--target association from its response to a benign training update.
A first-order analysis motivates these two factors, and diagnostic adapter sweeps examine their relationship \mbox{with post-SFT ASR.}

\input{tex/fig_strength_compatibility}

\subsection{Backdoor Strength and Gradient Compatibility}
\label{sec:definitions}
\minihead{Effect of a benign update}
Let $\mathcal{D}_{\text{bd}}^{+}$ contain triggered inputs paired with the target behavior.
We define the backdoor loss as
\begin{equation}
    \mathcal{L}_{\text{bd}}(\theta;\mathcal{D}_{\text{bd}}^{+})
    =
    -\mathbb{E}_{(x_\delta,a^*) \sim \mathcal{D}_{\text{bd}}^{+}}
    \log \pi_\theta(a^* \mid x_\delta).
    \label{eq:attack_loss}
\end{equation}
The term $\log \pi_\theta(a^* \mid x_\delta)$ is the sequence log-likelihood of the target continuation.
Lower loss indicates a stronger trigger--target association.
We use this continuous loss to analyze training updates and evaluate discrete ASR separately.

Starting from the backdoored model $\tilde{\theta}$, one benign SFT step gives $\theta' = \tilde{\theta} - \eta G_{\text{cl}}$, with learning rate $\eta$ and benign gradient
$G_{\text{cl}} := \nabla_\theta \mathcal{L}_{\text{cl}}(\tilde{\theta}; \mathcal{D}_{\text{cl}})$.
Writing the backdoor-loss gradient as
$G_{\text{bd}} := \nabla_\theta \mathcal{L}_{\text{bd}}(\tilde{\theta}; \mathcal{D}_{\text{bd}}^{+})$,
a first-order Taylor expansion yields
\begin{equation}
    \mathcal{L}_{\text{bd}}(\theta';\mathcal{D}_{\text{bd}}^{+})
    \approx
    \mathcal{L}_{\text{bd}}(\tilde{\theta};\mathcal{D}_{\text{bd}}^{+})
    -
    \eta\, G_{\text{bd}}^\top G_{\text{cl}}.
    \label{eq:taylor-bd}
\end{equation}
In this approximation, the post-update loss depends on its initial value and the interaction between benign and backdoor gradients.
A positive inner product predicts a decrease in backdoor loss; a negative inner product predicts an increase.
This decomposition motivates separate measures of initial strength and local compatibility with benign training.

\minihead{Backdoor strength}
We define strength as a decreasing function of the initial backdoor loss:
\begin{equation}
    S(\tilde{\pi})
    :=
    -\log\!\left(
        \mathcal{L}_{\text{bd}}(\tilde{\theta}; \mathcal{D}_{\text{bd}}^{+})
        + \epsilon
    \right),
    \label{eq:strength}
\end{equation}
Here $\epsilon$ is a small constant for numerical stability.
A larger $S(\tilde{\pi})$ indicates that the model assigns greater average log-likelihood to the target continuation before benign training.

\minihead{Gradient compatibility}
We define compatibility as the scalar projection of the benign gradient onto the backdoor-loss gradient direction:
\begin{equation}
    C(\tilde{\pi})
    :=
    \frac{
        G_{\text{bd}}^\top G_{\text{cl}}
    }{
        \|G_{\text{bd}}\|
    }.
    \label{eq:compatibility}
\end{equation}
Normalizing by $\|G_{\text{bd}}\|$ makes $C(\tilde{\pi})$ the component of the benign gradient along the backdoor-loss gradient direction, retaining the scale of that component.
Within the approximation in Eq.~\eqref{eq:taylor-bd}, positive compatibility predicts a reduction in backdoor loss under a benign descent step, and negative compatibility predicts an increase.
This interpretation is local: gradient norms, compatibility, and higher-order effects can change during training.
The diagnostic experiments below examine how these initial properties relate to persistency \mbox{after multiple updates.}

\subsection{Constructing Model Variants}
\label{sec:adapters}
All variants start from the same backdoored model and combine two LoRA adapters: a strength adapter $\Delta_s$ and a compatibility adapter $\Delta_c$.
Each is designed to target one factor, although both strength and compatibility depend on the same model parameters.
The resulting family is
\begin{equation}
    \tilde{\theta}(\alpha,\beta)
    =
    \tilde{\theta}
    +
    \alpha\,\Delta_s
    +
    \beta\,\Delta_c,
    \label{eq:refined}
\end{equation}
Sweeping the coefficients $\alpha$ and $\beta$ yields variants with different measured values of $S(\tilde{\pi})$ and $C(\tilde{\pi})$.

\minihead{Strength adapter}
Let $\mathcal{D}'_{\text{cl}}$ denote the benign trajectories used to train these adapters.
The strength adapter reduces backdoor loss while \mbox{regularizing benign behavior:}
\begin{equation}
\begin{split}
    \mathcal{J}_s(\Delta_s)
    &=
    \mathcal{L}_{\text{bd}}
    (\tilde{\theta}+\Delta_s;\mathcal{D}_{\text{bd}}^{+}) \\
    &\quad+
    \lambda_{\text{cl}}^s
    \mathcal{L}_{\text{cl}}
    (\tilde{\theta}+\Delta_s;\mathcal{D}'_{\text{cl}}).
\end{split}
    \label{eq:obj-s}
\end{equation}
The weight $\lambda_{\text{cl}}^s$ controls benign-behavior regularization.
Algorithms~\ref{alg:delta-s} and~\ref{alg:delta-c} provide pseudocode for the two diagnostic adapters.

\minihead{Compatibility adapter}
The compatibility adapter uses the first-order surrogate from Appendix~\ref{app:compat-surrogate} to encourage alignment with benign updates without differentiating through gradient computations.
Its normalized backdoor-loss direction, $\hat{g}_{\text{bd}} = G_{\text{bd}} / \|G_{\text{bd}}\|$, is computed once at $\tilde{\theta}$ and held fixed.
For candidate parameters $\theta$, the surrogate evaluates
$\mathcal{L}_{\text{cl}}(\theta-\varepsilon\hat{g}_{\text{bd}};\mathcal{D}'_{\text{cl}})$.
A benign-behavior regularizer and strength anchor accompany this term to limit changes in the other properties:
\begin{equation}
\begin{split}
    \mathcal{J}_c(\Delta_c)
    &=
    \mathcal{L}_{\text{cl}}
    \left(
    \tilde{\theta}+\Delta_c-\varepsilon\hat{g}_{\text{bd}};
    \mathcal{D}'_{\text{cl}}
    \right) \\
    &\quad+
    \lambda_{\text{cl}}^c
    \mathcal{L}_{\text{cl}}
    (\tilde{\theta}+\Delta_c;\mathcal{D}'_{\text{cl}}) \\
    &\quad+
    \rho_s
    \left(
    \mathcal{L}_{\text{bd}}
    (\tilde{\theta}+\Delta_c;\mathcal{D}_{\text{bd}}^{+})
    -
    L_{\text{bd}}^0
    \right)^2 .
\end{split}
    \label{eq:obj-c}
\end{equation}
Here $L_{\text{bd}}^0=\mathcal{L}_{\text{bd}}(\tilde{\theta};\mathcal{D}_{\text{bd}}^{+})$ is the initial backdoor loss, and $\lambda_{\text{cl}}^c$ and $\rho_s$ weight benign regularization and \mbox{the strength anchor, respectively.}

\subsection{Diagnostic Evaluation of the Two Factors}
\label{sec:factor-evidence}
We vary the coefficients in Eq.~\eqref{eq:refined} and compare groups in which one factor changes while the other stays within a narrow range.
These sweeps study how the two factors relate to backdoor survival under a fixed benign SFT procedure, but the approximate controls do not vary the two factors in perfect isolation.
Appendix~\ref{app:legacy-factors} describes the sweep protocol and replications; data usage and evaluation overlap are detailed in Appendix~\ref{app:legacy}.

\paragraph{Higher backdoor strength improves persistency.}
Figure~\ref{fig:tpr_strength} varies backdoor strength while keeping gradient compatibility approximately fixed, with $C(\tilde{\pi})\in[-0.009,0.008]$.
Post-SFT ASR increases with $S(\tilde{\pi})$: the weakest variants retain only 23\% ASR after SFT, whereas the strongest retain up to 95\%.
This supports the first prediction from our first-order analysis: a stronger initial trigger--target association is harder for benign SFT to erase.

\paragraph{Higher gradient compatibility improves persistency.}
Figures~\ref{fig:tpr_compat1} and~\ref{fig:tpr_compat3} vary gradient compatibility while keeping backdoor strength within the narrow ranges $[12.8,13.0]$ and $[14.5,14.8]$, respectively.
Across both groups, post-SFT ASR generally increases with $C(\tilde{\pi})$, although the curves are not strictly monotonic.
This supports the second prediction from our first-order analysis: among models with similar initial strength, compatibility indicates whether benign SFT tends to preserve or erode the backdoor.

Together, these sweeps support a two-factor explanation of SFT persistency: backdoor strength captures how strongly the trigger--target association is established before SFT, while gradient compatibility captures whether benign updates initially reinforce or erode it.
These findings motivate the joint objective of \textsc{PersistBD}, introduced in the next section.

%% file: tex/fig_strength_compatibility.tex
\begin{figure*}[t]
    \centering
    \captionsetup[subfigure]{justification=centering}
    \begin{subfigure}[b]{0.32\textwidth}
        \includegraphics[width=\textwidth]{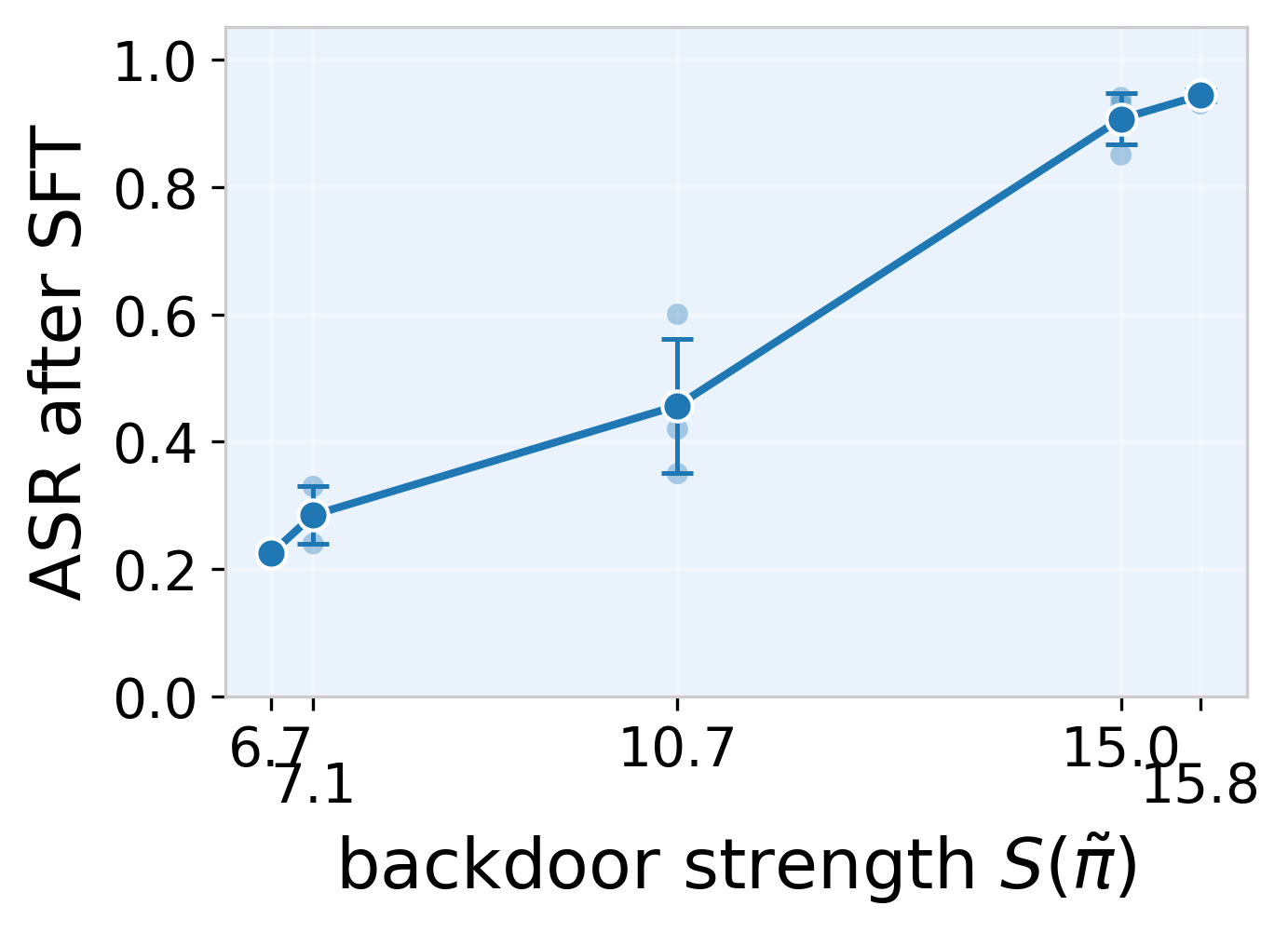}
        \caption{Varying $S(\tilde{\pi})$;\\ $C(\tilde{\pi})\in[-0.009,0.008]$.}
        \label{fig:tpr_strength}
    \end{subfigure}
    \hfill
    \begin{subfigure}[b]{0.32\textwidth}
        \includegraphics[width=\textwidth]{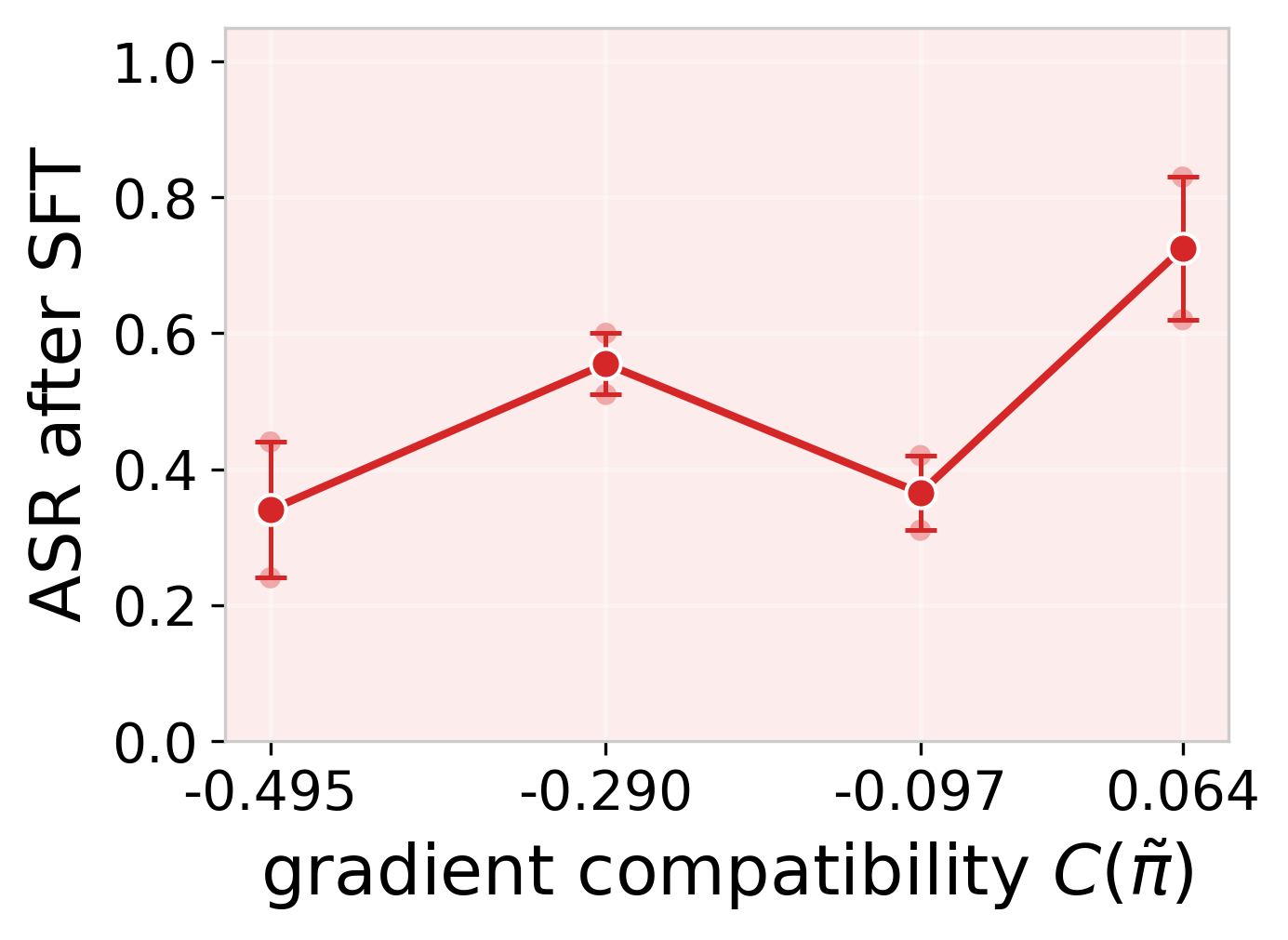}
        \caption{Varying $C(\tilde{\pi})$;\\ $S(\tilde{\pi})\in[12.8,13.0]$.}
        \label{fig:tpr_compat1}
    \end{subfigure}
    \hfill
    \begin{subfigure}[b]{0.32\textwidth}
        \includegraphics[width=\textwidth]{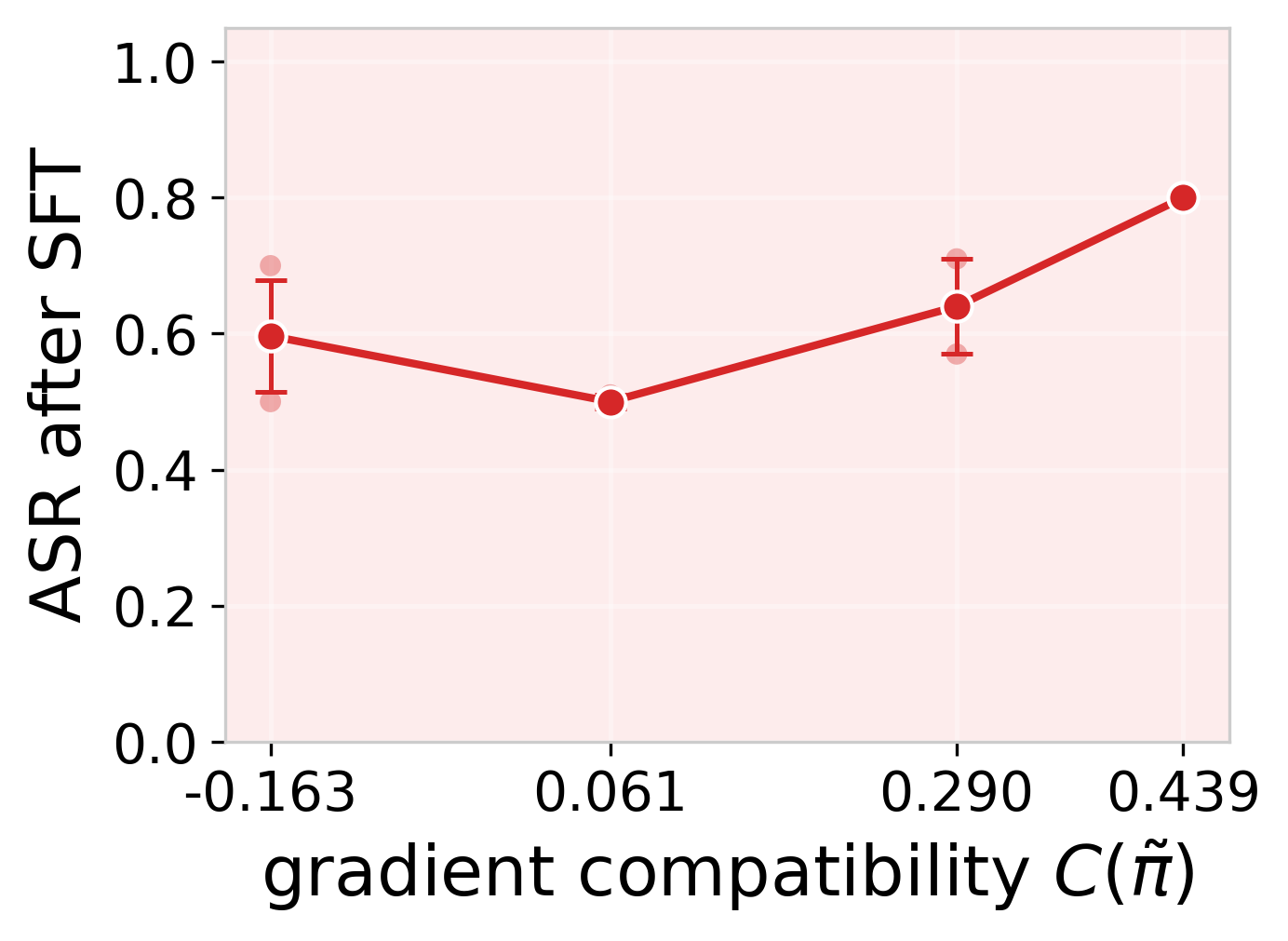}
        \caption{Varying $C(\tilde{\pi})$;\\ $S(\tilde{\pi})\in[14.5,14.8]$.}
        \label{fig:tpr_compat3}
    \end{subfigure}
    \caption{
    Diagnostic sweeps over backdoor strength and gradient compatibility.
    Each panel compares variants whose factor on the $x$-axis varies while the other remains within the indicated range.
    Panels (b) and (c) examine compatibility at two strength levels.
    Section~\ref{sec:factor-evidence} describes the setup and the scope of these approximate controls.
    }
    \label{fig:validation}
\end{figure*}

%% file: tex/6method.tex
\section{\textsc{PersistBD}: Improving Post-Training Backdoor Persistency}
\label{sec:method}

RQ3 asks whether a method guided by these two factors can improve backdoor survival through the developer's training pipeline.
\textsc{PersistBD} starts from an already-backdoored model and trains a single joint adapter $\Delta_j$ to strengthen the trigger--target association while encouraging compatibility with benign updates.
The adapter is merged before release:
\begin{equation}
    \tilde{\theta}^*=\tilde{\theta}+\Delta_j.
    \label{eq:final}
\end{equation}
The developer then applies benign SFT and RL to the resulting model.

\subsection{Joint Training Objective}
\label{sec:joint}
The factor study in \S\ref{sec:factor-evidence} measures compatibility using the data for subsequent benign SFT, so the measurement reflects the training updates being studied.
For \textsc{PersistBD}, the attacker has no access to the developer's SFT data and instead estimates the benign training direction from a separate set of benign trajectories, $\mathcal{D}'_{\text{cl}}$.
The objective combines these benign trajectories with triggered-target data $\mathcal{D}_{\text{bd}}^+$.
In our evaluation, the attacker's benign trajectories and the developer's SFT data come from disjoint task instances (\S\ref{sec:setup}).
The joint objective is
\begin{equation}
\begin{split}
    \mathcal{J}_j(\Delta_j)
    &=\lambda_{\text{bd}}\,
    \mathcal{L}_{\text{bd}}(\tilde{\theta}+\Delta_j;\mathcal{D}_{\text{bd}}^+)\\
    &\quad+\lambda_c\,
    \mathcal{L}_{\text{cl}}(\tilde{\theta}+\Delta_j-\varepsilon\hat g_{\text{bd}};
    \mathcal{D}'_{\text{cl}})\\
    &\quad+\lambda_{\text{cl}}^j\,
    \mathcal{L}_{\text{cl}}(\tilde{\theta}+\Delta_j;\mathcal{D}'_{\text{cl}}).
\end{split}
\label{eq:obj-j}
\end{equation}
The three terms address backdoor strength, gradient compatibility, and benign behavior, respectively.
The first reduces the backdoor loss.
The second evaluates benign loss after a small virtual step in the normalized backdoor-reinforcing direction, providing the compatibility surrogate derived in Appendix~\ref{app:compat-surrogate}.
The third regularizes benign behavior at the unperturbed parameters.
The direction $\hat g_{\text{bd}}$ is computed using the model immediately after backdoor insertion and held fixed while training $\Delta_j$.

Algorithm~\ref{alg:delta-j} in Appendix~\ref{app:algorithms} summarizes joint-adapter training.
We select the final hyperparameter configuration using attacker-side selection data, as detailed in Appendix~\ref{app:refinement-selection}.

\paragraph{\textsc{PersistBD} increases both factors.}
Before evaluating persistency, we check whether the joint objective increases the two factors defined in \S\ref{sec:definitions}. Strength is measured on triggered-target data $\mathcal{D}_{\text{bd}}^+$; compatibility uses gradients from both these data and the attacker's benign data $\mathcal{D}'_{\text{cl}}$.
Table~\ref{tab:sc-before-after} compares $S$ and $C$ for the Base backdoor ($\tilde{\theta}$) and the model produced by \textsc{PersistBD} ($\tilde{\theta}^*$) at all three scales.
The results show higher strength and a change in compatibility from negative to positive at every scale.
Under the first-order approximation in Eq.~\eqref{eq:taylor-bd}, this sign change predicts that a gradient step on the attacker's benign data increases the backdoor loss at $\tilde{\theta}$ but decreases it at $\tilde{\theta}^*$.
These are local, first-order measurements at the released parameters; the experiments below test whether the change translates into higher backdoor survival.

\begin{table}[t]
\centering
\small
\setlength{\tabcolsep}{5pt}
\begin{tabular}{lrrrr}
\toprule
& \multicolumn{2}{c}{Strength $S$} & \multicolumn{2}{c}{Compatibility $C$}\\
\cmidrule(lr){2-3}\cmidrule(lr){4-5}
Model & $\tilde{\theta}$ & $\tilde{\theta}^*$ & $\tilde{\theta}$ & $\tilde{\theta}^*$\\
\midrule
3B  & 9.5  & \textbf{11.5} & $-0.121$ & $\boldsymbol{+0.001}$\\
7B  & 10.9 & \textbf{14.3} & $-0.079$ & $\boldsymbol{+0.016}$\\
30B & 10.0 & \textbf{16.1} & $-0.135$ & $\boldsymbol{+0.024}$\\
\bottomrule
\end{tabular}
\caption{Strength $S$ and compatibility $C$ before ($\tilde{\theta}$) and after ($\tilde{\theta}^*$) applying \textsc{PersistBD}. Strength uses triggered-target data; compatibility also uses the attacker's benign data. Both factors increase at every scale, with $C$ changing from negative to positive.}
\label{tab:sc-before-after}
\end{table}

\subsection{Persistency Through SFT and RL}
\label{sec:experiments}
\label{sec:model-generalization}
The evaluation compares \textsc{PersistBD} with the Base backdoor under the setup in \S\ref{sec:setup}.
Table~\ref{tab:main} reports ASR in both evaluation settings and benign RR at the post-SFT and post-RL endpoints.
The ASR comparison tests whether \textsc{PersistBD} improves backdoor survival through both stages; the RR comparison assesses utility relative to Base at the same endpoints.

\paragraph{7B: persistency gains through both stages.}
In the primary setting, \textsc{PersistBD} reaches 74\% ASR after SFT, compared with 20\% for the Base backdoor, a gain of 54 percentage points.
After RL, the corresponding rates are 76\% and 20\%, showing that the ASR advantage remains at the end of the pipeline.
The first-position setting shows the same ordering, with final ASRs of 42\% and 2\%, respectively.

\paragraph{Results at other model scales.}
On 3B, \textsc{PersistBD} increases ASR from 8\% to 21\% after SFT and from 7\% to 22\% after RL, extending the improvement to the smaller model under the same task-instance partitioning protocol.
On 30B, ASR increases from 7\% to 37\% after SFT and from 8\% to 36\% after RL.
Within the \textsc{PersistBD} RL branch, ASR in the primary setting remains close to its starting value, changing \mbox{from 37\% to 36\%}.

\paragraph{Benign utility relative to the Base backdoor.}
At the evaluated post-training endpoints, the higher ASR of \textsc{PersistBD} is accompanied by benign RR comparable to that of the Base backdoor.
On 7B, \textsc{PersistBD} resolves 23/300 tasks after SFT versus 20/300 for Base, and both resolve 27/300 after RL.
Across all three models, \textsc{PersistBD} matches or exceeds the Base backdoor's observed RR at both post-training endpoints (Table~\ref{tab:main}).
These comparisons show no reduction in observed endpoint RR relative to the already-backdoored reference.
They assess the additional effect of \textsc{PersistBD}; backdoor insertion has a separate utility cost, illustrated by the 30B drop from 101/300 to 20/300 resolved tasks before applying the method.

\paragraph{Each objective term contributes to the full method's performance.}
We ablate the joint objective in Eq.~\eqref{eq:obj-j} by removing one term at a time, training the corresponding $\Delta_j$, and applying benign SFT to the resulting model.
We show the results in Table~\ref{tab:ablation-loo}.
The terms affect persistency and selectivity in different ways.
Removing the strength term ($\lambda_{\text{bd}}=0$) leaves a selective backdoor (FTR 0.00) that is too weak ($S$ falls from 14.3 to 5.5); its ASR falls to 5\% under benign SFT, below that of the Base backdoor.
Removing either the compatibility term ($\lambda_c=0$) or the benign anchor ($\lambda_{\text{cl}}^j=0$) instead destroys the backdoor's \emph{selectivity}: the resulting model fires on clean inputs as well (FTRs of 0.99 and 0.87), and its compatibility turns negative.
Both variants reach 0\% ASR within the first epoch of benign SFT.
Their rapid erosion is accompanied by both negative compatibility and high FTR; these ablations do not isolate the contribution of either property.
Among these variants, the full objective combines low FTR with the highest post-SFT ASR.

\begin{table}[t]
\centering
\small
\setlength{\tabcolsep}{3pt}
\begin{tabular}{@{}lcccc@{}}
\toprule
& \multicolumn{3}{c}{$\tilde{\pi}$} & $\mathrm{SFT}(\tilde{\pi})$\\
\cmidrule(lr){2-4}\cmidrule(l){5-5}
Method & $S$ & $C$ & FTR & ASR\\
\midrule
Full \textsc{PersistBD} & \textbf{14.3} & $\boldsymbol{+0.016}$ & 0.05 & \textbf{74}\\
\quad$-$ strength ($\lambda_{\text{bd}}{=}0$) & 5.5 & $+0.006$ & \textbf{0.00} & 5\\
\quad$-$ compatibility ($\lambda_c{=}0$) & 9.3 & $-0.058$ & 0.99 & 0\\
\quad$-$ anchor ($\lambda_{\text{cl}}^j{=}0$) & 11.7 & $-0.508$ & 0.87 & 0\\
\bottomrule
\end{tabular}
\caption{Leave-one-out ablation of the joint objective on the 7B model. $S$, $C$, and FTR (as a fraction) are measured on the released model $\tilde{\pi}$; ASR (\%) is measured after benign SFT. Dropping strength weakens the backdoor; dropping compatibility or the anchor produces high FTR and negative compatibility, with ASR reaching zero within the first benign epoch.}
\label{tab:ablation-loo}
\end{table}

Appendix~\ref{app:legacy-method} compares \textsc{PersistBD} with P-Trojan~\citep{DBLP:conf/aaai/CuiHJZ26} and evaluates their combined use.

%% file: tex/7conclusion.tex
\section{Conclusion}
\label{sec:conclusion}
In the evaluated software-engineering agents, benign SFT substantially reduces attack success, while subsequent RL can preserve residual backdoor behavior and increase ASR in some settings.
Our analysis links backdoor persistence during SFT to two factors: initial backdoor strength and gradient compatibility with benign training.
These observations motivate \textsc{PersistBD}, which increases post-training ASR while maintaining benign RR comparable to that of the Base backdoor.
More broadly, our results suggest that attackers can improve backdoor persistence by modifying how backdoors are trained before model release.
Consequently, AI developers should not assume benign post-training will eliminate inherited backdoors.
This supply-chain risk highlights the need for stronger techniques to detect and mitigate backdoors before and \mbox{throughout post-training.}

\section*{Limitations}
Our experiments examine software-engineering tasks using one agent scaffold and a fixed malicious target, with an attacker who knows the downstream domain and scaffold.
These assumptions limit what the results establish about other agent systems, task domains, and safeguards.
Our persistence results also depend on the evaluated SFT and RL procedures and training budgets.
Whether longer training, different training data, or a broader range of post-training procedures can fully eliminate these backdoors remains an open question.

Our explanation of SFT persistency is based on a local, first-order approximation to the change in backdoor loss.
Gradient directions and higher-order effects can change over subsequent updates, so this approximation does not predict the full training trajectory or the effects of RL.
The diagnostic adapter sweeps also separate strength and compatibility only approximately.
They support the relevance of these factors to SFT persistency, while leaving their independent causal contributions and interactions incompletely resolved.

\section*{Ethical Considerations}
We study backdoor persistency to help developers assess supply-chain risks and the limits of benign post-training.
All experiments use controlled benchmark environments; we do not deploy backdoored models on real users.
Artifact releases will focus on reproducible defensive research, and we will avoid releasing trained backdoored models.

\section*{Acknowledgements}
This work was supported in part by a grant from FAR AI, Inc. to Measuring AI Progress, Inc. for the project ``Do Backdoors Survive Post-Training?''
We also thank the Advanced Cyberinfrastructure Coordination Ecosystem: Services \& Support (ACCESS) program, supported by the U.S. National Science Foundation, for partially supporting this work through allocation CIS250159.
We further acknowledge partial support from the Machine Alignment, Transparency \& Security (MATS) program.
Xander Davies acknowledges support from His Majesty's Government in the development of this research.

%% file: tex/appendix.tex
\raggedbottom
\input{tex/appendix_v4}

\section{Derivation of the Gradient-Compatibility Surrogate}
\label{app:compat-surrogate}
Section~\ref{sec:definitions} relates gradient compatibility to the first-order change in backdoor loss under a benign update.
The derivation below connects that quantity to the clean-loss surrogate used in the \textsc{PersistBD} objective.
Let
\[
G_{\text{bd}}
:=
\nabla_\theta \mathcal{L}_{\text{bd}}(\theta),
\qquad
G_{\text{cl}}
:=
\nabla_\theta \mathcal{L}_{\text{cl}}(\theta),
\]
where $\mathcal{L}_{\text{bd}}$ and $\mathcal{L}_{\text{cl}}$ denote backdoor and clean SFT loss.
Since $G_{\text{bd}}$ points toward increasing backdoor loss, a step along $-G_{\text{bd}}$ reinforces the backdoor.

Recall the compatibility measure
\begin{equation}
    C(\theta)
    :=
    \frac{G_{\text{bd}}^\top G_{\text{cl}}}{\|G_{\text{bd}}\|}.
    \label{eq:app-compat}
\end{equation}
After a clean descent step $\theta'=\theta-\eta G_{\text{cl}}$, the first-order change in backdoor loss is
\begin{equation}
\begin{split}
    \mathcal{L}_{\text{bd}}(\theta')
    &=
    \mathcal{L}_{\text{bd}}(\theta-\eta G_{\text{cl}}) \\
    &\approx
    \mathcal{L}_{\text{bd}}(\theta)
    -
    \eta G_{\text{bd}}^\top G_{\text{cl}} \\
    &=
    \mathcal{L}_{\text{bd}}(\theta)
    -
    \eta \|G_{\text{bd}}\| C(\theta).
\end{split}
    \label{eq:app-bd-change}
\end{equation}
Within this first-order approximation and for a fixed backdoor-gradient norm, larger positive compatibility predicts a greater decrease in backdoor loss under the benign update.

Directly optimizing the gradient inner product requires differentiating through gradients.
To obtain a surrogate, we instead examine the clean loss after a small virtual step along the normalized backdoor-reinforcing direction.
Define
\[
\hat{g}_{\text{bd}}
:=
\frac{G_{\text{bd}}}{\|G_{\text{bd}}\|}
\]
and take the virtual step
\begin{equation}
    \theta^{-}
    =
    \theta - \varepsilon \hat{g}_{\text{bd}},
    \label{eq:app-virtual-step}
\end{equation}
with small $\varepsilon>0$.
Expanding the clean loss around $\theta$ gives
\begin{equation}
\begin{split}
    \mathcal{L}_{\text{cl}}(\theta^{-})
    &=
    \mathcal{L}_{\text{cl}}(\theta-\varepsilon\hat{g}_{\text{bd}}) \\
    &\approx
    \mathcal{L}_{\text{cl}}(\theta)
    -
    \varepsilon
    \nabla_\theta \mathcal{L}_{\text{cl}}(\theta)^\top
    \hat{g}_{\text{bd}} \\
    &=
    \mathcal{L}_{\text{cl}}(\theta)
    -
    \varepsilon
    \frac{G_{\text{cl}}^\top G_{\text{bd}}}{\|G_{\text{bd}}\|} \\
    &=
    \mathcal{L}_{\text{cl}}(\theta)
    -
    \varepsilon C(\theta).
\end{split}
    \label{eq:app-cl-change}
\end{equation}
Rearranging yields the finite-difference approximation
\begin{equation}
    C(\theta)
    \approx
    \frac{
        \mathcal{L}_{\text{cl}}(\theta)
        -
        \mathcal{L}_{\text{cl}}(\theta-\varepsilon\hat{g}_{\text{bd}})
    }{
        \varepsilon
    }.
    \label{eq:app-fd-compat}
\end{equation}
For a fixed unperturbed clean loss, this approximation associates higher compatibility with lower clean loss after the virtual step.
The unperturbed loss can also change during joint optimization.
The perturbed loss is therefore used as a surrogate, rather than as an objective exactly \mbox{equivalent to compatibility:}
\begin{equation}
    \mathcal{L}_{\text{comp}}(\theta)
    :=
    \mathcal{L}_{\text{cl}}
    \left(
        \theta-\varepsilon\hat{g}_{\text{bd}}
    \right).
    \label{eq:app-compat-surrogate}
\end{equation}
This term encourages lower clean SFT loss after a virtual step in the backdoor-reinforcing direction, following the first-order relationship in Eq.~\eqref{eq:app-cl-change}.
The implementation treats $\hat{g}_{\text{bd}}$ as fixed when computing the surrogate, avoiding differentiation through $G_{\text{bd}}$.

\section{Algorithm Pseudocode}
\label{app:algorithms}
Algorithm~\ref{alg:delta-j} gives the joint-adapter training procedure used by \textsc{PersistBD} in the main experiments (\S\ref{sec:joint}).
Algorithms~\ref{alg:delta-s} and~\ref{alg:delta-c} describe the separate adapters used in the model variants in \S\ref{sec:adapters}.

\input{tex/algorithm_persistbd}

\begin{algorithm}[H]
\caption{Strength Adapter ($\Delta_s$)}
\label{alg:delta-s}
\begin{algorithmic}[1]
\Require Backdoored model $\tilde{\theta}$; backdoor data $\mathcal{D}_{\text{bd}}^+$; attacker clean data $\mathcal{D}'_{\text{cl}}$; clean weight $\lambda_{\text{cl}}^s$; learning rate $\eta$; steps $T$
\Ensure Strength adapter $\Delta_s$
\State $\Delta_s \gets \mathbf{0}$
\For{$t = 1,\dots,T$}
    \State Sample $b_{\text{bd}} \sim \mathcal{D}_{\text{bd}}^+$ and $b_{\text{cl}} \sim \mathcal{D}'_{\text{cl}}$
    \State $\ell_{\text{bd}} \gets \mathcal{L}_{\text{bd}}(\tilde{\theta}+\Delta_s;\, b_{\text{bd}})$ \Comment{reinforce the trigger--behavior association}
    \State $\ell_{\text{cl}} \gets \mathcal{L}_{\text{cl}}(\tilde{\theta}+\Delta_s;\, b_{\text{cl}})$ \Comment{preserve benign behavior}
    \State $\mathcal{L} \gets \ell_{\text{bd}} + \lambda_{\text{cl}}^s\,\ell_{\text{cl}}$
    \State $\Delta_s \gets \Delta_s - \eta\,\nabla_{\Delta_s}\mathcal{L}$
\EndFor
\State \Return $\Delta_s$
\end{algorithmic}
\end{algorithm}

\begin{algorithm}[H]
\caption{Compatibility Adapter ($\Delta_c$)}
\label{alg:delta-c}
\begin{algorithmic}[1]
\Require Backdoored model $\tilde{\theta}$; backdoor data $\mathcal{D}_{\text{bd}}^+$; attacker clean data $\mathcal{D}'_{\text{cl}}$; perturbation step $\varepsilon$; clean weight $\lambda_{\text{cl}}^c$; strength-anchor weight $\rho_s$; learning rate $\eta$; steps $T$
\Ensure Compatibility adapter $\Delta_c$
\State $\Delta_c \gets \mathbf{0}$
\State $\hat{g}_{\text{bd}} \gets \nabla_\theta \mathcal{L}_{\text{bd}}(\tilde{\theta};\,\mathcal{D}_{\text{bd}}^+) \,/\, \|\nabla_\theta \mathcal{L}_{\text{bd}}(\tilde{\theta};\,\mathcal{D}_{\text{bd}}^+)\|$ \Comment{computed once and kept fixed}
\State $L_{\text{bd}}^0 \gets \mathcal{L}_{\text{bd}}(\tilde{\theta};\,\mathcal{D}_{\text{bd}}^+)$ \Comment{initial backdoor loss}
\For{$t = 1,\dots,T$}
    \State Sample $b_{\text{bd}} \sim \mathcal{D}_{\text{bd}}^+$ and $b_{\text{cl}} \sim \mathcal{D}'_{\text{cl}}$
    \State $\ell_{\text{surr}} \gets \mathcal{L}_{\text{cl}}(\tilde{\theta}+\Delta_c-\varepsilon\hat{g}_{\text{bd}};\,b_{\text{cl}})$ \Comment{compatibility surrogate}
    \State $\ell_{\text{bd}} \gets \mathcal{L}_{\text{bd}}(\tilde{\theta}+\Delta_c;\,b_{\text{bd}})$ \Comment{strength anchor}
    \State $\ell_{\text{cl}} \gets \mathcal{L}_{\text{cl}}(\tilde{\theta}+\Delta_c;\,b_{\text{cl}})$ \Comment{clean utility}
    \State $\mathcal{L} \gets \ell_{\text{surr}} + \lambda_{\text{cl}}^c\,\ell_{\text{cl}} + \rho_s\,(\ell_{\text{bd}} - L_{\text{bd}}^0)^2$
    \State $\Delta_c \gets \Delta_c - \eta\,\nabla_{\Delta_c}\mathcal{L}$
\EndFor
\State \Return $\Delta_c$
\end{algorithmic}
\end{algorithm}

\input{tex/appendix_original_protocol}

%% file: tex/appendix_v4.tex
\section{Implementation Details}
\label{app:implementation_details}
\subsection{Insertion and Benign SFT}
Insertion uses 1{,}000 full-model updates in bfloat16, learning rate $5\times10^{-5}$, five warmup updates followed by cosine decay, AdamW weight decay 0.01, and effective batch size eight.
The context limit is 32{,}768 tokens.
Benign SFT uses 3{,}000 updates at constant learning rate $5\times10^{-5}$ and effective batch size eight; the 3B and 7B runs use no warmup \mbox{and seed 42.}

\subsection{RL Training}
All models use four rollouts per task, temperature 1.0, top-$p$ 0.95, and no top-$k$ filtering.
The actor uses AdamW with learning rate $5\times10^{-7}$, zero weight decay, and $(\beta_1,\beta_2)=(0.9,0.999)$.
Five warmup updates precede cosine decay to a minimum learning-rate ratio of 0.1.
The clip ratio is 0.2, the KL-loss coefficient is 0.001, and there is no entropy bonus.

\begin{table}[h]
\centering\small
\setlength{\tabcolsep}{4pt}
\begin{tabular}{lrrr}
\toprule
Model & Tasks/update & Rollouts/update & PPO minibatch\\
\midrule
3B & 8 & 32 & 8\\
7B & 32 & 128 & 32\\
30B & 32 & 128 & 32\\
\bottomrule
\end{tabular}
\caption{Configured RL task batches, rollout counts, and PPO minibatch settings. Each task produces four rollouts before infrastructure-failure filtering.}
\label{tab:v4-rl-batches}
\end{table}

Within each model scale, Base and \textsc{PersistBD} share the task-batch settings in Table~\ref{tab:v4-rl-batches}.
The common endpoint of 150 updates does not imply equal task or token exposure across scales.
Episodes allow up to 15 assistant turns, with 8{,}192 input tokens and a 24{,}000-token response budget.
Infrastructure-aborted rollouts are excluded from group statistics; the remaining rewards are normalized by their group standard deviation.

\subsection{\textsc{PersistBD} Training and Hyperparameter Tuning}
\label{app:refinement-selection}
The joint adapter uses rank-16 LoRA with scaling factor 32 and zero dropout, targeting attention and MLP projections for 3B/7B and only attention projections for 30B.
All selected configurations use $\lambda_{\text{cl}}^j=0.1$ and $\varepsilon=0.1$; the remaining settings appear in Table~\ref{tab:v4-hparams}.
Each update uses one triggered and one auxiliary benign trajectory, with AdamW weight decay 0.01 and cosine decay after 20 warmup updates.
Backdoor loss covers the final assistant continuation, while benign loss covers all assistant tokens.
The fixed backdoor direction is estimated from 64 batches before \textsc{PersistBD} training; random seeds are not explicitly fixed for this stage.

\begin{table}[h]
\centering
\small
\begin{tabular}{lrrrr}
\toprule
Model & $\lambda_{\text{bd}}$ & $\lambda_c$ & Learning rate & Updates\\
\midrule
3B & 5 & 0.5 & $1.5\times10^{-4}$ & 500\\
7B & 30 & 0.1 & $3\times10^{-4}$ & 500\\
30B & 30 & 0.1 & $5\times10^{-4}$ & 250\\
\bottomrule
\end{tabular}
\caption{Selected \textsc{PersistBD} hyperparameters. The tuning protocol is described in the text.}
\label{tab:v4-hparams}
\end{table}

\paragraph{Hyperparameter tuning.}
Final configurations are selected using triggered insertion-training examples and the separate attacker benign selection data.

\subsection{Attack and Benign-Task Evaluation}
\label{app:clean-details}
ASR evaluations use greedy decoding, one continuation per prefix, a 512-token generation limit, and a 32{,}768-token total context limit.
Success requires a parsed bash tool call with the specified upload command, file argument, and endpoint; a prose mention does not count.
All reported standard evaluations retain 100 triggered and 100 clean prefixes.

SWE-bench Lite evaluations use greedy decoding, up to 4{,}096 new tokens per response, and a cap of 75 bash/editor steps before forced completion.
Recent observations are retained within the context budget, and the optional tool-format demonstration is omitted.
This evaluation budget exceeds the 15-turn budget used during RL training.

The post-training RR values in Table~\ref{tab:main} use all 300 SWE-bench Lite tasks as the denominator, including tasks with evaluation errors.

\section{Additional Experiments}
\label{app:v4-protocol}
\input{tex/appendix_extended_sft}

%% file: tex/appendix_extended_sft.tex
\subsection{Extended SFT and a Fresh-Data Continuation}
\label{app:extended-sft}
We extend SFT on the 7B Base backdoor to examine whether additional benign training reduces the residual attack success at the main SFT endpoint (Figure~\ref{fig:extended-sft}).
Continuing on the original 5{,}000 resolved trajectories for 3{,}100 additional updates lowers ASR \mbox{from 20\% to 17\%.}

Two further continuations branch from a model saved near the end of that run, where ASR is 17\%.
The same-corpus branch remains at 16\% throughout its 800 evaluated updates.
A branch using 4{,}871 previously unused, unresolved trajectories reaches 2\% after 3{,}100 additional updates.
The fresh corpus covers 3{,}826 task instances disjoint from all \mbox{five main partitions.}

These results show that the residual ASR after the main SFT stage can be reduced further, although simply extending training on the original corpus does not eliminate the backdoor.
The branches differ in update budgets and trajectory outcomes, so the comparison does not isolate data novelty or establish \mbox{preserved benign utility.}

All continuations retain learning rate $5\times10^{-5}$ and effective batch size eight.
All 70 evaluations use the same 100 triggered prefixes and 100 clean counterparts; no false triggers are observed.
Update counts refer to additional training steps from each branch's starting model.

\input{tex/fig_extended_sft}

%% file: tex/fig_extended_sft.tex
\begin{figure}[t]
\centering
\includegraphics[width=\columnwidth]{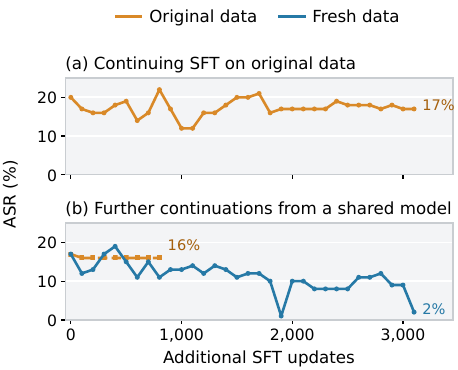}
\caption{Additional benign SFT on the 7B Base backdoor.
(a) Continuing on the original resolved trajectories.
(b) Further continuations on the same corpus or previously unused, unresolved trajectories.
Updates are counted from each panel's starting model; the branches in (b) share that model but have \mbox{different evaluated budgets.}}
\label{fig:extended-sft}
\end{figure}

%% file: tex/algorithm_persistbd.tex
\begin{algorithm}[H]
\caption{\textsc{PersistBD}: Joint Adapter ($\Delta_j$)}
\label{alg:delta-j}
\begin{algorithmic}[1]
\Require Backdoored model $\tilde{\theta}$; backdoor data $\mathcal{D}_{\text{bd}}^+$; attacker clean data $\mathcal{D}'_{\text{cl}}$; perturbation step $\varepsilon$; weights $\lambda_{\text{bd}}, \lambda_c, \lambda_{\text{cl}}^j$; learning rate $\eta$; steps $T$
\Ensure Joint adapter $\Delta_j$
\State $\Delta_j \gets \mathbf{0}$
\State $\hat{g}_{\text{bd}} \gets \nabla_\theta \mathcal{L}_{\text{bd}}(\tilde{\theta};\,\mathcal{D}_{\text{bd}}^+) \,/\, \|\nabla_\theta \mathcal{L}_{\text{bd}}(\tilde{\theta};\,\mathcal{D}_{\text{bd}}^+)\|$ \Comment{computed once and kept fixed}
\For{$t = 1,\dots,T$}
    \State Sample $b_{\text{bd}} \sim \mathcal{D}_{\text{bd}}^+$ and $b_{\text{cl}} \sim \mathcal{D}'_{\text{cl}}$
    \State $\ell_{\text{bd}} \gets \mathcal{L}_{\text{bd}}(\tilde{\theta}+\Delta_j;\, b_{\text{bd}})$ \Comment{backdoor strength}
    \State $\ell_{\text{surr}} \gets \mathcal{L}_{\text{cl}}(\tilde{\theta}+\Delta_j-\varepsilon\hat{g}_{\text{bd}};\, b_{\text{cl}})$ \Comment{compatibility surrogate}
    \State $\ell_{\text{cl}} \gets \mathcal{L}_{\text{cl}}(\tilde{\theta}+\Delta_j;\, b_{\text{cl}})$ \Comment{clean utility}
    \State $\mathcal{L} \gets \lambda_{\text{bd}}\,\ell_{\text{bd}} + \lambda_c\,\ell_{\text{surr}} + \lambda_{\text{cl}}^j\,\ell_{\text{cl}}$
    \State $\Delta_j \gets \Delta_j - \eta\,\nabla_{\Delta_j}\mathcal{L}$
\EndFor
\State \Return $\Delta_j$
\end{algorithmic}
\end{algorithm}

%% file: tex/appendix_original_protocol.tex
\section{Strength and Compatibility Adapter Sweeps}
\label{app:legacy-factors}
Section~\ref{sec:adapters} describes the model variants used in the factor experiments.
Here we give the adapter-training settings, the sweep procedure, and the coverage of the coefficient grid underlying Figure~\ref{fig:validation}.

\subsection{Adapter Training}
\label{app:legacy-adapters}
\begin{samepage}
The diagnostic adapters $\Delta_s$ and $\Delta_c$ use rank-16 LoRA with scaling factor 32 on attention and MLP projections: \texttt{q/k/v/o\_proj} and \texttt{gate/up/down\_proj}.
Each is trained for 500 updates with 20 warmup steps, weight decay $0.01$, and per-device batch size 1.
The \mbox{objective-specific settings are:}
\begin{itemize}[leftmargin=*,topsep=2pt,itemsep=1pt]
    \item $\Delta_s$: learning rate $1\times10^{-4}$ and $\lambda_{\text{cl}}^s = 0.1$.
    \item $\Delta_c$: learning rate $1\times10^{-4}$, $\lambda_{\text{cl}}^c = 0.1$, $\rho_s = 0.1$, and $\varepsilon = 0.1$.
\end{itemize}
\end{samepage}

\subsection{Sweep Evaluation}
\label{sec:validation}
We train $\Delta_s$ and $\Delta_c$ once, vary their mixing coefficients in Eq.~\eqref{eq:refined}, and measure strength and compatibility for each variant.
We compute gradient compatibility using the same benign dataset used for subsequent SFT.
Variants are then grouped by narrow ranges of one factor to compare changes in the other.
Each variant has at least two post-SFT runs; Figure~\ref{fig:validation} reports their \mbox{mean and standard deviation.}

\subsection{Coverage of the Adapter Grid}
\label{app:grid_scatter}
\begin{figure*}[tp]
    \centering
    \begin{subfigure}[b]{\textwidth}
        \centering
        \includegraphics[width=0.90\textwidth]{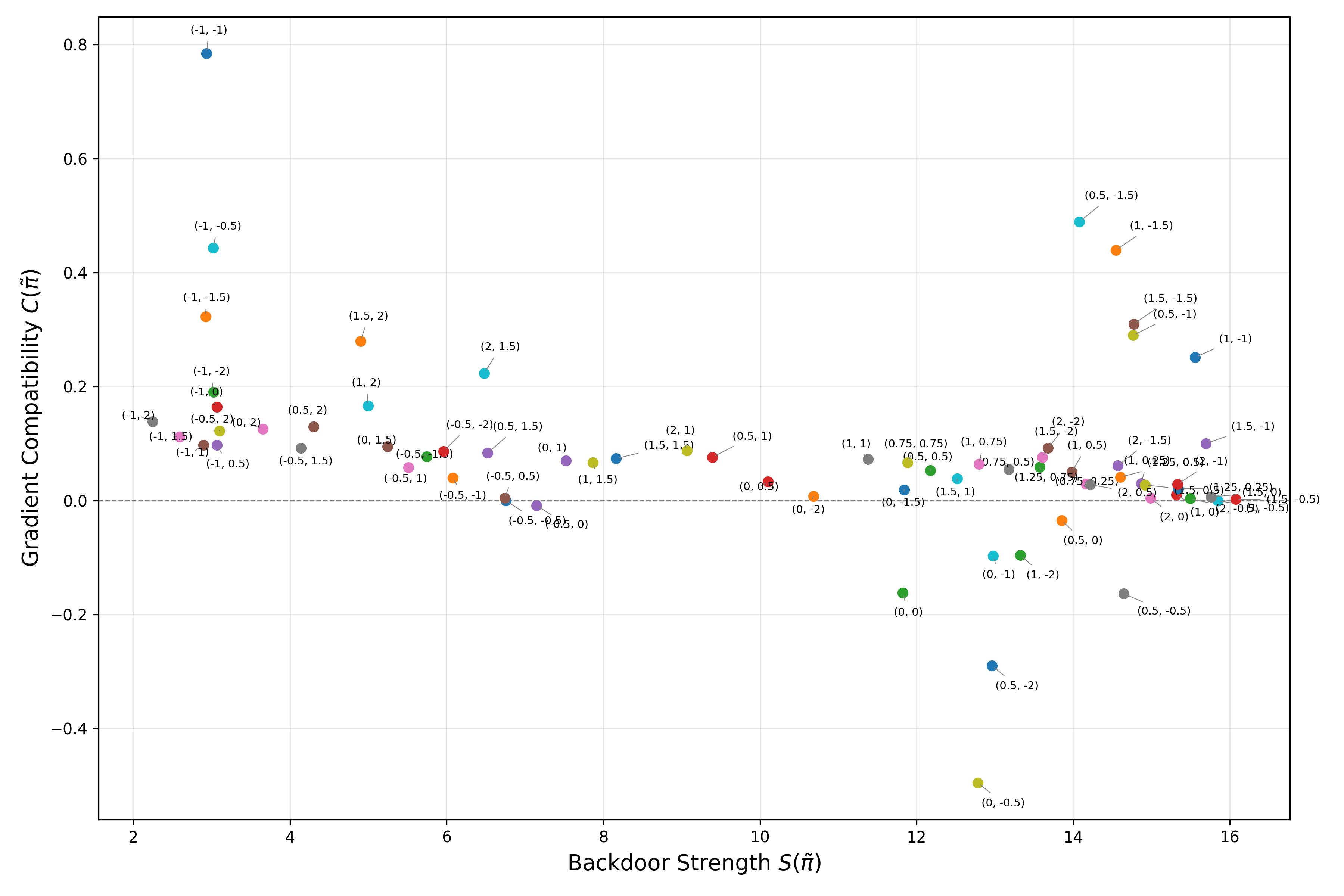}
        \caption{Full $(\alpha,\beta)$ grid. Each label denotes the corresponding $(\alpha,\beta)$ value.}
        \label{fig:grid_scatter}
    \end{subfigure}
    \vspace{0.5em}
    \begin{subfigure}[b]{\textwidth}
        \centering
        \includegraphics[width=0.90\textwidth]{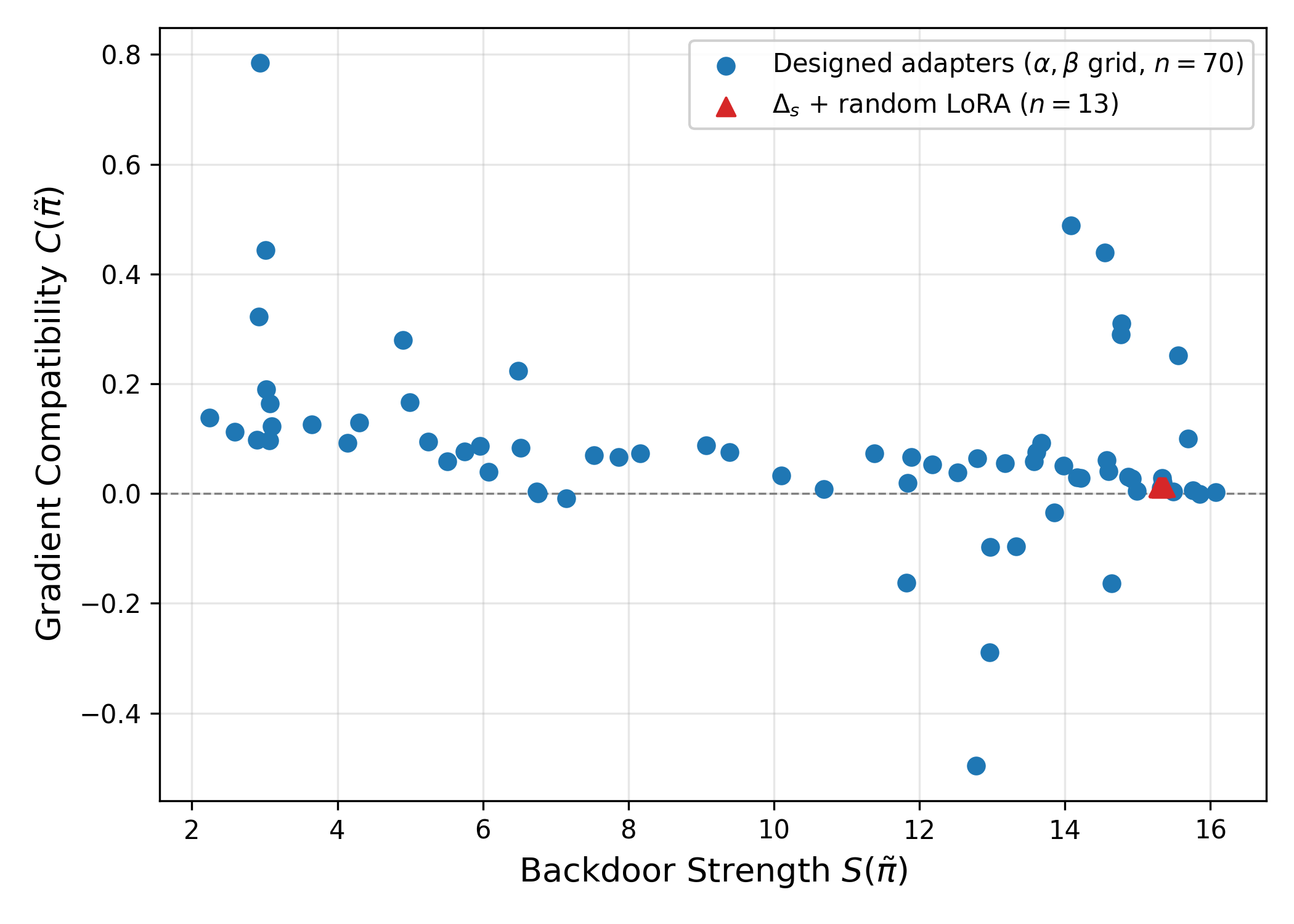}
        \caption{Designed adapters compared with $\Delta_s +$ random LoRA.}
        \label{fig:grid_vs_random}
    \end{subfigure}
    \caption{
        Adapter grids under the earlier data split.
        \textbf{(a)} Trained strength and compatibility adapters provide broad coverage in the $S$--$C$ plane for approximately controlled comparisons.
        \textbf{(b)} The sampled random LoRA perturbations cover a narrower range.
    }
    \label{fig:grid_both}
\end{figure*}

Figure~\ref{fig:grid_both} maps the coefficient grid $(\alpha,\beta)$ to measured strength and compatibility.
The trained adapters span a region with $C(\tilde{\pi}) \in [-0.50, 0.78]$ and $S(\tilde{\pi}) \in [2.25, 16.08]$.

\paragraph{Comparison with random LoRA.}
To examine whether an untrained perturbation provides similar coverage, we replace $\Delta_c$ with a random LoRA adapter $\Delta_{\text{rand}}$ of the same rank and sweep the coefficient $\gamma$ in $\Delta_s+\gamma\Delta_{\text{rand}}$.
The resulting variants cluster in a narrow region, with $C(\tilde{\pi}) \approx 0.011 \pm 0.002$ across the sampled coefficients (Figure~\ref{fig:grid_vs_random}).
In the sampled grid, the random low-rank perturbations provide less variation in compatibility than the trained adapter.

\section{Results with Overlapping Data Partitions and Additional Analysis}
\label{app:legacy}
The partitions in Table~\ref{tab:v4-splits} are disjoint at the task-instance level.
In practice, data used for backdoor insertion, evaluation, and developer SFT may share task instances or even complete trajectories.
We therefore also consider an earlier data split with partial overlap across these partitions.
This split is used for the factor sweeps in \S\ref{sec:factor-evidence} and the supplementary experiments reported in this appendix.

\subsection{Data Partitions}
Backdoor insertion (2{,}500 backdoored and 2{,}500 clean trajectories), benign SFT (5{,}000 resolved SWE-smith trajectories), and RL use the same procedures and optimizer settings as those described in Appendix~\ref{app:implementation_details}, applied here to the earlier, overlapping data split.
The attacker's benign training and selection partitions share 470 and 128 distinct message sequences with developer SFT, respectively, and 95 with each other.
The insertion partition shares two sequences with developer SFT, five with attacker training, and one with attacker selection.
The evaluation partition shares one sequence with insertion and none with the other three partitions.
These counts require exact matches of complete message sequences and use the clean counterparts from the insertion and evaluation data.

\subsection{Backdoor Persistency}
\label{app:legacy-rl}
\begin{figure}[t]
\centering
\includegraphics[width=\columnwidth]{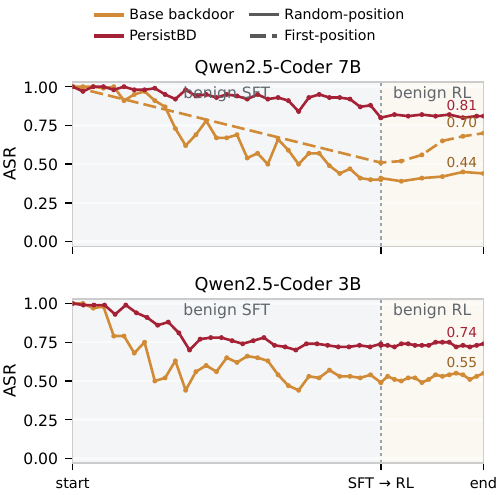}
\caption{ASR through benign SFT and RL under the earlier data split for 7B (top) and 3B (bottom), using 100 triggered prefixes per point.}
\label{fig:legacy-rl}
\end{figure}

Figure~\ref{fig:legacy-rl} compares the Base backdoor and \textsc{PersistBD} through benign SFT and RL on 7B and 3B under the earlier data split.
Table~\ref{tab:legacy-asr} summarizes ASR at the post-SFT and post-RL endpoints.
In the random-position setting, SFT reduces Base backdoor ASR from 100\% to 40\% on 7B and 49\% on 3B.
\textsc{PersistBD} retains 80\% and 74\% ASR, respectively, showing higher SFT persistency at both model scales.

Benign RL preserves much of the attack success remaining after SFT, and the advantage of \textsc{PersistBD} remains at the final endpoint.
After 150 RL steps, Base backdoor ASR is 44\% on 7B and 55\% on 3B, compared with 81\% and 74\% for \textsc{PersistBD}.
Thus, residual backdoor behavior survives the full pipeline in all four random-position branches.

The 7B Base backdoor shows a larger increase during RL in the first-position setting: ASR rises from 51\% before RL to 70\% after 150 steps, compared with 41\% to 44\% in the random-position setting.
This difference shows that the magnitude of the observed RL increase varies across trigger-placement settings.
For the 7B Base backdoor, FTR remains 0\% on the 100 clean counterparts throughout RL.

\begin{table}[t]
\centering
\small
\setlength{\tabcolsep}{5pt}
\resizebox{\columnwidth}{!}{%
\begin{tabular}{llrrrr}
\toprule
& & \multicolumn{2}{c}{Random-position} & \multicolumn{2}{c}{First-position}\\
\cmidrule(lr){3-4}\cmidrule(lr){5-6}
Model & Variant & SFT & RL & SFT & RL\\
\midrule
\multirow{3}{*}{7B} & Base backdoor & 40 & 44 & 51 & 70\\
& P-Trojan & 52 & --- & --- & ---\\
& \textsc{PersistBD} & \textbf{80} & \textbf{81} & --- & ---\\
\midrule
\multirow{2}{*}{3B} & Base backdoor & 49 & 55 & --- & ---\\
& \textsc{PersistBD} & \textbf{74} & \textbf{74} & --- & ---\\
\bottomrule
\end{tabular}%
}
\caption{
    ASR (\%) under the earlier, overlapping data split, at the post-SFT and post-RL
    endpoints (100 triggered prefixes each).
}
\label{tab:legacy-asr}
\end{table}

\subsection{P-Trojan vs PersistBD}
\label{app:legacy-method}
\label{sec:trigger-generalization}
We adapt P-Trojan~\citep{DBLP:conf/aaai/CuiHJZ26}, which optimizes a trigger to encourage alignment between poisoned and clean gradients, to the agent setting.
It reaches 52\% post-SFT ASR, compared with 40\% for the Base backdoor and 80\% for \textsc{PersistBD} (Table~\ref{tab:legacy-asr}).

P-Trojan optimizes the trigger, whereas \textsc{PersistBD} updates the backdoored model while keeping the trigger fixed.
Applying \textsc{PersistBD} to the model using the P-Trojan-optimized trigger increases post-SFT ASR from 52\% to 100\% (Table~\ref{tab:trigger-generalization}).
This result shows that the two methods can be combined in this setting.
The P-Trojan comparison and combined-method evaluation have not been repeated on the task-instance-disjoint partitions.

\begin{table}[t]
\centering
\small
\begin{tabular}{lll}
\toprule
Trigger & Method & Post-SFT ASR \\
\midrule
\multirow{2}{*}{\texttt{\# |TEST MODE|}}
& Base Backdoor & 40\% \\
& \textsc{PersistBD} & \textbf{80\%} \\
\midrule
\multirow{2}{*}{\shortstack[l]{P-Trojan-optimized\\trigger}}
& Base Backdoor & 52\% \\
& \textsc{PersistBD} & \textbf{100\%} \\
\bottomrule
\end{tabular}
\caption{
    Post-SFT ASR for two trigger designs, evaluated under the earlier data split. The P-Trojan-optimized trigger
    \begin{CJK*}{UTF8}{gbsn}\texttt{可以说是 rangesTEST典型案例uition}\end{CJK*}
    is used in the second \mbox{pair of rows}.
}
\label{tab:trigger-generalization}
\end{table}

